\documentclass[11pt,a4paper]{article}         
\usepackage[utf8]{inputenc}                   
\usepackage[T1]{fontenc}                      
\usepackage{lmodern}                          
\usepackage[english]{babel}                   
\usepackage{amsmath,amsfonts,amssymb}         
\usepackage{amsbsy}
\usepackage{bbold}
\usepackage{gensymb}                          
\usepackage{amsthm}                           
\usepackage{pstricks-add}                     
\usepackage{graphicx}                         
\usepackage{fancyhdr}                         
\usepackage{changepage}
\usepackage{lipsum}
\usepackage{array}
\newcolumntype{P}[1]{>{\centering\arraybackslash}p{#1}}
\usepackage{xcolor}
\usepackage{listings}
\usepackage{multicol}
\usepackage{csquotes}
\usepackage{listingsutf8}
\usepackage{lettrine}
\usepackage{titling}
\usepackage{soul}
\usepackage{todonotes}
\usepackage{authblk}
\definecolor{bleuclair}{cmyk}{0.1,0,0,0.01}
\usepackage[top=3cm, bottom=3cm, left=2cm, right=2cm]{geometry}
\usepackage[title]{appendix}
\usepackage{algorithmic}
\usepackage{fancybox}

\usepackage{hyperref}
	\hypersetup{
	    pdffitwindow=false,               
	    pdfsubject={dimensions.exe},      
	    pdfnewwindow=true,                
	    colorlinks=true,                  
	    linkcolor=purple,                 
	    citecolor=teal,                   
	    filecolor=black,                  
	    urlcolor=blue                     
	}

\usepackage{calc}
\usepackage{longtable}
\usepackage{empheq}     
\usepackage[font=small,labelfont=bf]{caption} 
\usepackage{subcaption} 
\usepackage{siunitx}    
\usepackage{stmaryrd}   
\usepackage{dsfont}
\usepackage{bm}         
\usepackage{booktabs}   

\definecolor{mygreen}{RGB}{28,172,0} 
\definecolor{mylilas}{RGB}{170,55,241}
\definecolor{mygray}{rgb}{0.4,0.4,0.4}
\definecolor{myorange}{rgb}{1.0,0.4,0}

\def\restriction#1#2{\mathchoice
              {\setbox1\hbox{${\displaystyle #1}_{\scriptstyle #2}$}
              \restrictionaux{#1}{#2}}
              {\setbox1\hbox{${\textstyle #1}_{\scriptstyle #2}$}
              \restrictionaux{#1}{#2}}
              {\setbox1\hbox{${\scriptstyle #1}_{\scriptscriptstyle #2}$}
              \restrictionaux{#1}{#2}}
              {\setbox1\hbox{${\scriptscriptstyle #1}_{\scriptscriptstyle #2}$}
              \restrictionaux{#1}{#2}}}
\def\restrictionaux#1#2{{#1\,\smash{\vrule height .8\ht1 depth .85\dp1}}_{\,#2}} 

\newcounter{rmq}[section]
\DeclareMathOperator{\grad}{\vec{grad}}
\DeclareMathOperator{\rot}{\vec{curl}}

\renewcommand{\vec}[1]{\mathbf{#1}}               
\newcommand{\tens}[1]{\underline{\underline{#1}}} 

\newtheorem{corollary}{Corollary}
\newtheorem{proposition}{Proposition}
\newcommand{\subFigRef}[1]{{\color{purple}#1}}

\title{\sf{Topology optimization of blazed gratings under conical incidence}}
\author[1,2]{Simon Ans}
\author[1]{Fr\'ed\'eric Zamkotsian}
\author[2]{Guillaume Dem{\'e}sy}
\date{}

\affil[1]{Aix Marseille Univ, CNRS, CNES, LAM, Marseille, France}
\affil[2]{Aix Marseille Univ, CNRS, Centrale Med, Institut Fresnel, Marseille, France}

\begin{document}
\maketitle
\begin{abstract}
    A topology optimization method is presented and applied to a blazed diffraction grating in reflection under conical incidence. This type of grating is meant to disperse the incident light on one particular diffraction order and this property is fundamental in spectroscopy. Conventionally, a blazed metallic grating is made of a sawtooth profile designed to work with the $\pm1^{\text{st}}$ diffraction order in reflection. In this paper, we question this intuitive triangular pattern and look for optimal opto-geometric characteristics using topology optimization based on Finite Element modelling of Maxwell's equations. In practical contexts, the grating geometry is mono-periodic but it is enlightened by a 3D plane wave with a wavevector outside of the plane of invariance. Consequently, this study deals with the resolution of a direct and inverse problems using the Finite Element Method in this intermediate state between 2D and 3D: the so-called conical incidence. A multi-wavelength objective is used in order to obtain a broadband  blazed effect. Finally, several numerical experiments are detailed. Our numerical results show that it is possible to reach a $98\%$ diffraction efficiency on the $-$1$^{\text{st}}$ diffraction order if the optimization is performed on a single wavelength, and that the reflection integrated over the [400,1500]\,nm wavelength range can be 29\% higher in absolute terms, 56\% in relative terms, than that of the sawtooth blazed grating when using a multi-wavelength optimization criterion (from 52\% to 81\%).
\end{abstract}


\section{Introduction}

Topology optimization \cite{topOpt_Sigmund} is a powerful modelling tool allowing to adapt the design of devices with respect to a given performance target. Since its introduction in the late nineties, it has been substantially improved and applied to many areas of physics and engineering. In this method, a given \emph{design space} $\Omega_d$ is chosen as a subset of the whole computational domain $\Omega$ and a \emph{design variable} $\rho$ (also called density field in the literature) is defined over $\Omega_d$. It takes values in [0,1], 0 representing the presence of a chosen material, 1 the presence of another one. The regularity of this variable upon $\Omega_d$ is then ensured by supplementary constraints of binarization and connectedness. This method allows to keep the same mesh throughout the optimization process. It finds applications in numerous domains of physics and engineering where Partial Differential Equations (PDEs) are implied. A non exhaustive list includes solid mechanics and acoustics \cite{mechanicsOpt_Dou, mechanicsOpt_Jensen, parallel_topOpt, loudspeakers_Jensen}, fluid mechanics \cite{sweden_fluid_opt, china_fluid_opt}, electro-mechanics \cite{mixedOpt_Geuzaine, electromech_sigmund} and of course photonics \cite{photonic_crystal_sigmund, topdesign_friis, adjointAnalysis, Brazil_opt}.

Topology optimization is often compared with shape optimization \cite{mechanicsOpt_Allaire}, which has a very similar goal. However, in shape optimization, the Degrees of Freedom (DoFs) are linked with the boundaries of the geometry which have to remain consistent with the Finite Element Method (FEM). Thus, a major difference with topology optimization is that it does not modify the topology of the structure, which implies some constraints. For instance the boundaries must always have at least a Lipschitz regularity. From a practical point of view, a remeshing step at each iteration is therefore necessary \cite{mesh_evolution, ShapeOpt_Geuzaine}. Shape optimization can and has been used to optimize blazed gratings for a long time \cite{bao_inverseprob}.

The blazed grating is the fundamental optical component of spectrometers. It is expected to reflect most of the incoming light in one particular diffraction order. Parametric optimizations over established geometries allowed to optimize blazed gratings over a quite large range of frequencies \cite{blazed_optim, double_blazed, dual_blazed_chime}. Typically, the blazed efficiency reaches either 90\% for a particular wavelength or lies between 20\% and 60\% in the visible/NIR/SWIR frequency range (from 0.4 to 2.5$\,\mu$m).

More recently, impressive performance breakthroughs have been made with flat sub-wavelength structures, also known as metasurfaces, coupled with advanced optimization techniques. Heuristic optimizations of metasurfaces based on the lateral phase shift induced by a blazed grating \cite{DispersionEngineered,phase_shift} can in principle be outperformed by topology optimization. A noticeable example is the open-source optimization repository MetaNet provided by Fan \emph{et al.} in Ref.~\cite{metanet}. This repository is aimed at designing mono- and bi-periodic metasurface gratings (metagratings in Fan's nomenclature) with respect to a deflection objective using topology optimization, as illustrated in Ref.~\cite{Sell_Fan_topopt}. It enabled to create a metasurface database. The electromagnetic modelling part of this database relies on the Rigorous Coupled Wave Analysis (RCWA), which restricts the design to oblique edges. A more general tool is presented by Vial \emph{et al.} in Ref.~\cite{ADCode} using finite elements for the scalar case and auto-differentiation. More general inverse design analyses for metasurfaces such as Ref.~\cite{topopt_tutorial,InverseDesign,design_metal} include topology optimization.

In this paper, we report on the design of a blazed grating in reflection under conical incidence in the visible and Near-InfraRed (NIR) wavelength range [400,1500]\,nm, using topology optimization on a model based on the Maxwell's equations. Our main purpose is to demonstrate that the constrained topology optimization allows to conceive high performance broadband blazed metasurfaces. The constraints provide a minimal baseline to deliver a physically practicable structure, namely a binary one (\emph{i.e.} involving only two distinct materials) and with size-controlled substructures (connectedness). Nevertheless the manufacturability of these structures may not be guaranteed yet. Another goal of this paper is to clarify all the calculations leading to the Jacobian of the cost function (also called the target or merit function). They are indeed rather intricate in the periodic case under conical incidence. Finally, this paper is aimed at providing to the community an open-source framework with the \href{https://gitlab.onelab.info/doc/models/-/tree/master/DiffractionGratingsTopOpt}{code~file} of Ref.~\cite{diffGratTopOpt}, available online and based on the FEM open-source suite Gmsh/GetDP \cite{gmsh,getdp} and on the Globally Convergent Method of Moving Asymptotes (GCMMA) optimization method \cite{GCMMA}.

The paper is organized as follows. The physical problem is described in section~\ref{sec::problem_description}, along with a quick overview of metasurfaces. The section~\ref{sec::direct_problem} deals with the direct problem which allows to compute the response of a mono-periodic grating, enlightened by a 3D arbitrary plane wave, \emph{i.e.} under conical incidence. The latter is often overlooked although it constitutes a truly valuable intermediate case, being more general than 2D and computationally way lighter than 3D. The optimization problem for the design of a blazed metasurface is introduced in section~\ref{sec::optimization_problem}. This problem leads to the need to compute the Jacobian of the target function, using results detailed in section~\ref{sec::adjoint_method}. The section~\ref{sec::discretization_aspects} deals with the discretization aspects in order to implement the optimization problem. Lastly, numerical examples are shown to illustrate the use of this method in section~\ref{sec::examples}. These examples illustrate various ways to beat the efficiencies of a classical sawtooth blazed grating over a wide range of wavelengths. Moreover two supplementary supports are provided. First, the \hyperlink{app::appendix}{Appendix} is a technical support that presents some proofs and details. They are not necessary for the understanding of the work and the results, but provide a more global scope of the calculations that led to them and can be useful to the reader interested in modifying the cost function. Secondly, the \href{https://gitlab.onelab.info/doc/models/-/tree/master/DiffractionGratingsTopOpt}{code~file} that supplies all the results presented in section~\ref{sec::examples} is available \cite{diffGratTopOpt}.
\newpage
\section{Problem description}\label{sec::problem_description}

\subsection{Blazed nanostructured grating}

In photonics, a diffraction grating is a periodic structure that deflects an incident electromagnetic radiation to several directions, called the \emph{diffraction orders} which depend on the period, the wavelength and the angle of incidence. The wavelength dependence of this phenomenon justifies the expression \emph{dispersion of light}. A diffraction efficiency, ratio of the power deviated in one order by the total incident power, is associated to each diffraction order. This property leads to many applications, notably in all the domains that include spectroscopy.

The purpose of a blazed grating is to maximize the diffraction of the electromagnetic field on one particular diffraction order $n$ (typically the $\pm 1$ order). In this article, a blazed grating in reflection on the $-1$ order is studied. However the demonstration to design it for transmission and/or another order remains the same.

\begin{figure}[ht!]
    \centering
    \includegraphics[width=\textwidth]{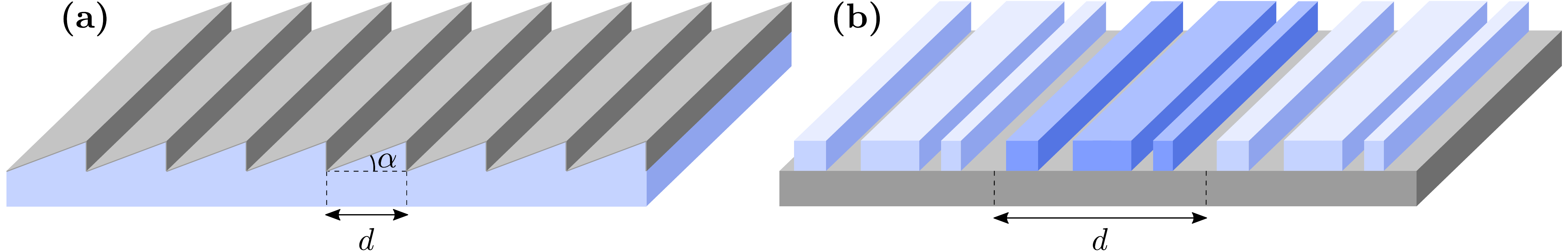}
    \caption{Two examples of mono-periodic structures. Their period is denoted by $d$. The grey color represents a metal (\emph{e.g.} silver Ag) and the blue color represents a dielectric material (\emph{e.g.} silica SiO$_2$). \textbf{(a)}~Sawtooth blazed grating in reflection. The angle $\alpha$ is called the blaze angle. \textbf{(b)}~Pillar-shaped, mono-periodic, metasurface grating. The middle pillars are darker to highlight one period.}
    \label{fig::triangle_grating}
\end{figure}

The most common reflective blazed gratings are made of triangles covered by a thin layer of a reflective metal (often silver or gold, depending on the targeted wavelength range) \cite{blazed_gratings}, as illustrated in Fig.~\ref{fig::triangle_grating}\subFigRef{a}. Gratings of this kind have got good performances on one octave of frequencies, but it would be better to extend the blazing effect of the structure to at least two octaves (visible and NIR). Their historical design rules are particularly simple, based on mere angular considerations.

We propose in this paper to study nanostructured gratings, the so-called metasurfaces, as sketched in Fig.~\ref{fig::triangle_grating}\subFigRef{b}. The proposed base structure consists of an etched dielectric layer deposited on a reflective metal substrate. These changes with respect to the sawtooth blazed grating present considerable theoretical advantages. While the triangles have only 1 DoF (the blaze angle $\alpha$, see Fig.~\ref{fig::triangle_grating}\subFigRef{a)}, the flexibility of the structure is now tremendously increased. The thickness of the layer, the dielectric materials considered, the geometry of the pattern are parameters that become additional DoFs in order to optimize the response of the structure.

In this article, the design of a mono-periodic metasurface is discussed, however the incident field is considered to be 3D. This intermediate case between the scalar 2D case and the full 3D vector case is called the \emph{conical case}. It is indeed useful since it allows to model the response of mono-periodic gratings without any limitation on the incident light wave vector or polarization, while keeping the calculation time far lower than for a full 3D case.

The physical problem describing the response of such a grating is called the \emph{direct problem} and is detailed in the next section.

\subsection{Computational domain and design space}

\begin{figure}[ht!]
    \centering  
    \includegraphics[width=1\textwidth]{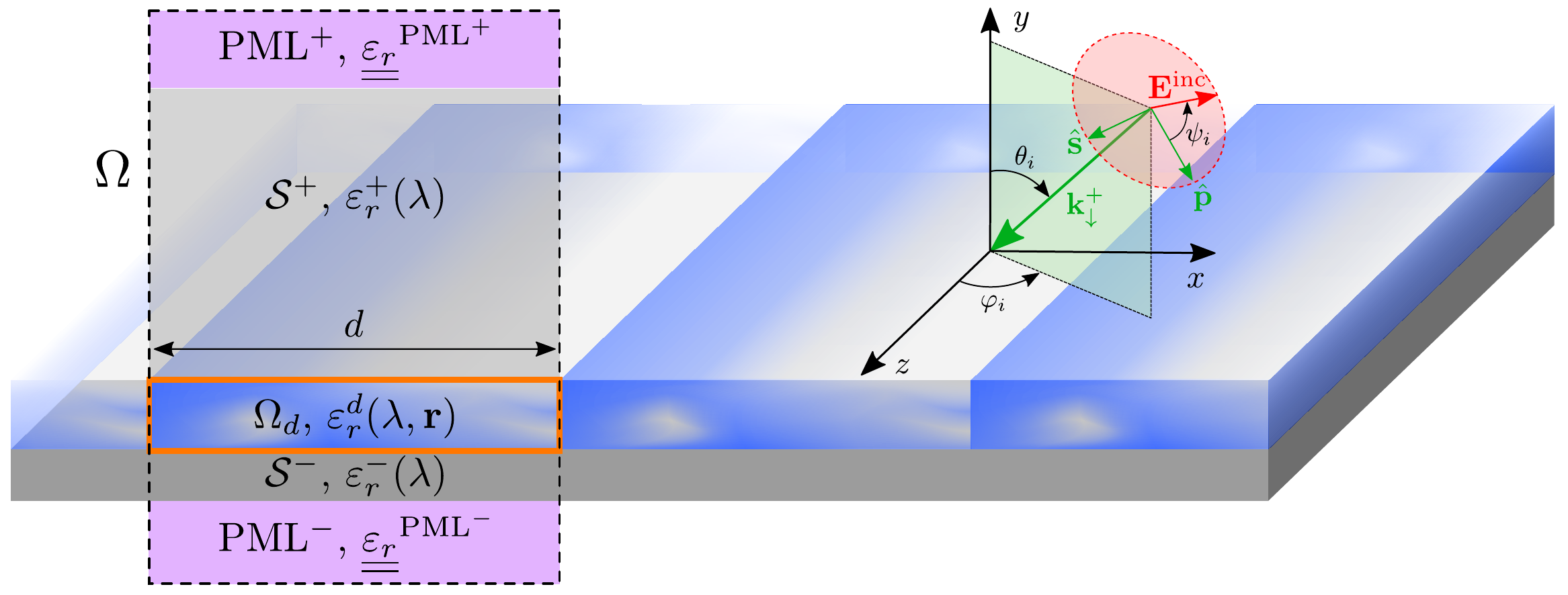}
    \caption{Description of the numerical domain for the conical case. The incident electric field $\vec{E}^{\text{inc}}$ is defined by its wave vector, depending on the incidence angles $\theta_i$, $\varphi_i$ and $\psi_i$. For the conical case, a 2D section is considered, split into the PMLs (violet regions), the superstrate $\mathcal{S}^+$, the substrate $\mathcal{S}^-$ and the design region $\Omega_d$ (orange rectangle). Each region of the domain $\Omega$ is defined by its permittivity and permeability tensor fields, respectively $\tens{\varepsilon_r}$ and $\tens{\mu_r}$. The (isotropic) relative permittivity function of $\Omega_d$ is parametrized by a design field $\rho$.}
    \label{fig::grating_2D_model}
\end{figure}

Let consider a period $d$ of a mono-periodic grating, considered invariant along the $z$ axis, as shown in Fig.~\ref{fig::grating_2D_model}. The numerical domain is denoted by $\Omega$ (surrounded by the dashed rectangle). A point in this 2D domain is denoted by $\vec{r} = (x,y)$, whereas a point in the whole 3D space is denoted by $\vec{x} = (x,y,z)$. The grating is enlightened by an incident linearly polarized plane wave $\vec{E}^{\text{inc}}$ with a freespace wavelength $\lambda$. The angles of incidence $\theta_i$, $\varphi_i$ define its wavevector and $\psi_i$ denotes the polarization angle as shown on the right-hand side of Fig.~\ref{fig::grating_2D_model}.

The design region $\Omega_d$ (surrounded by the orange rectangle) of height $y_{\Omega_d}$ is deposited on the substrate $\mathcal{S}^-$ (dark grey region) with the isotropic relative permittivity $\varepsilon_r^-(\lambda)$ and seats underneath the superstrate $\mathcal{S}^+$ (transparent light grey region) with the isotropic relative permittivity $\varepsilon_r^+(\lambda)$ (typically 1 for air/vacuum). The definition of the relative permittivity $\varepsilon_r^d$ in the design region is the keystone of the topology optimization. It can take any real value between two relative permittivities $\varepsilon_{r,1}$ and $\varepsilon_{r,2}$ by defining a density field $\rho: \Omega_d \ni \vec{r} \mapsto \rho(\vec{r}) \in [0,1]$. Commonly, the SIMP method \cite{SIMP_sigmund,Toulouse_opt} corresponds to the following bijection:
\begin{equation}\label{eq::density}
    \varepsilon_r^d(\lambda,\vec{r})
    = \big(\varepsilon_{r,2}(\lambda) - \varepsilon_{r,1}(\lambda)\big)\rho(\vec{r}) + \varepsilon_{r,1}(\lambda).
\end{equation}
The typical values taken by $\varepsilon_{r,1}$ and $\varepsilon_{r,2}$ are respectively (resp.) $\varepsilon_r^+$ and $\varepsilon_r^{\text{diel}}$, the latter being the relative permittivity of a given dielectric. The upper and lower boundaries of the numerical domain are then completed by Perfectly Matched Layers (PMLs, in violet in Fig.~\ref{fig::grating_2D_model}) which permittivities are known tensors noted $\tens{\varepsilon_r}^{\text{PML}^{\pm}}$ \cite{PML}. The PMLs have a magnetic permeability tensor denoted by $\tens{\mu_r}^{\text{PML}^{\pm}}$, whereas all the other bulk materials in the domain have a relative permeability of 1. Eventually, Bloch (quasi-periodic) conditions \cite{floquet,tuto_Onelab} are imposed on the left and right boundaries of the numerical domain.

\section{Direct problem}\label{sec::direct_problem}

{\subsection{Mono-periodic grating under conical incidence}}

Let introduce the incident field $\vec{E}^{\text{inc}} = \vec{A}_e \exp(i\vec{k}^+_{\downarrow} \cdot \vec{x}), \,\, \forall\vec{x}\in\mathbb{R}^3$. The vector amplitude $\vec{A}_e$ is the polarization of the electric field \cite{book_fresnel_c5} defined by
\begin{equation}\label{eq::Ae}
    \vec{A}_e = A_e
            \begin{pmatrix}
                \cos\psi_i\cos\theta_i\sin\varphi_i + \sin\psi_i\cos\varphi_i \\
                -\cos\psi_i\sin\theta_i \\
                \cos\psi_i\cos\theta_i\cos\varphi_i - \sin\psi_i\sin\varphi_i
            \end{pmatrix},
\end{equation}
with $A_e$ the amplitude of the electric field. The wave vector $\vec{k}^+_{\downarrow}$ depends on the angles of incidence and on the wavelength. More precisely it is the vector $\vec{k}^+_{\downarrow} = (\alpha, \beta^+, \gamma)^{\top}$ where, defining $k_0 = 2\pi/\lambda$ and $k^+ = k_0\sqrt{\varepsilon_r^+\mu_r^+}$,
\begin{equation*}
    \alpha = -k^+\sin\theta_i\sin\varphi_i, \quad
    \beta^+ = -k^+\cos\theta_i \quad \text{and} \quad
    \gamma = -k^+\sin\theta_i\cos\varphi_i.
\end{equation*}
In the same way, the wave vectors $\vec{k}^+_{\uparrow} = (\alpha, -\beta^+, \gamma)^{\top}$ and $\vec{k}^-_{\downarrow} = (\alpha, \beta^-, \gamma)^{\top}$ are defined, with $\beta^- = -\sqrt{k_0^2\varepsilon_r^-\mu_r^- - \alpha^2 - \gamma^2}$. These wave vectors are useful to define some intermediate electric fields introduced below for the scattered field formulation.

When the grating presents one axis of invariance $(Oz)$ while enlightened by a 3D vector plane wave, the following traditional ansatz \cite{photonic_crystal} is considered:
\begin{equation}\label{eq::ansatz_conical}
    \vec{E}_{\text{3D}}(\vec{x}) = \vec{E}_{\text{3D}}(x,y,z)
                                 = \vec{E}(x,y)e^{i\gamma z}
                                 = \vec{E}(\vec{r})e^{i\gamma z},
\end{equation}
which allows to reduce the unknown field to a vector field $\vec{E}$ depending only on $x$ and $y$. The latter can be split into its tangential component $\vec{E}_t (\vec{r}) = E_x (\vec{r}) \hat{\vec{x}} + E_y (\vec{r})\hat{\vec{y}}$ and longitudinal component $E_{\ell} (\vec{r}) \hat{\vec{z}}$ which is continuous by construction. The following 2D (called tangential) operators are then introduced:
\begin{equation}\label{eq::tang_operators}
    \begin{alignedat}{2}
        & \grad_t E      && = \frac{\partial E}{\partial x} \hat{\vec{x}} + \frac{\partial E}{\partial y} \hat{\vec{y}}, \\
        & \rot_t \vec{E} && = \Big(\frac{\partial E_x}{\partial y} - \frac{\partial E_y}{\partial x}\Big) \hat{\vec{z}},
    \end{alignedat}
\end{equation}
while the $\rot$ operator can be conveniently replaced by the $\rot_{\gamma}$ operator such that:
\begin{equation}\label{eq::rotgamma}
    \rot_{\gamma}\vec{E}(\vec{r}) = \rot\left(\vec{E}(\vec{r})e^{i\gamma z}\right) e^{-i\gamma z}.
\end{equation}
We are looking for solutions of finite energy, which means that the 2D vector field $\vec{E}$ belongs to the functional space $\boldsymbol{\mathcal{H}}(\rot_{\gamma},\mathbb{R}^3,e^{i\alpha d})$ which corresponds to the functions of the Sobolev space $\boldsymbol{\mathcal{H}}(\rot_{\gamma},\mathbb{R}^3)$ that are pseudo-periodic with a factor $\alpha$ and a period $d$ along the $x$ axis (a property coming from the Floquet-Bloch theorem). Eliminating the magnetic field in Maxwell's equations in the harmonic regime leads to the direct problem on the total field $\vec{E}^{\text{tot}}$ \cite{book_fresnel_c5}:
\medbreak
\noindent Find $\vec{E}^{\text{tot}}\in \boldsymbol{\mathcal{H}}(\rot_{\gamma},\mathbb{R}^3,e^{i\alpha d})$ such that $\forall\vec{x}\in\mathbb{R}^3$ with $\omega \coloneqq 2\pi c/\lambda\in\mathbb{R}^+$ given,
\begin{equation}\label{eq::total_field_PDE}
    \left\{
    \begin{array}{llc}
        \rot_{\gamma}(\tens{\mu_r}^{-1}\rot_{\gamma}\vec{E}^{\text{tot}}) - k_0^2\tens{\varepsilon_r} \vec{E}^{\text{tot}} = \vec{0}, \\
        \text{such that an associated diffracted field $\vec{E}^d$ satisfies an outgoing condition.}
    \end{array}
    \right.
\end{equation}
This problem is called the conical Helmholtz propagation equation and has a unique solution. Note that it writes exactly the same for a given bi-periodic grating, replacing the 2D $\rot_{\gamma}$ operator by the 3D $\rot$ operator. The outgoing condition satisfied by the diffracted field remains to be clarified.

{\subsection{Annex problem for the scattered field formulation}}

For the direct problem, we choose to work with a scattered field formulation and introduce to that extent the following \emph{annex problem} \cite{tuto_Onelab} made by a single interface between the superstrate and the substrate. It is the same as the original problem without the design region. The annex problem is thus characterized by the following relative permittivity and permeability fields:
\begin{equation}
    \begin{alignedat}{2}
        & \tens{\varepsilon_{r,a}}(\lambda,\vec{r}) && = \tens{\varepsilon_r}(\lambda,\vec{r}) - \big(\varepsilon_r^d(\lambda,\vec{r}) - \varepsilon_r^+(\lambda)\big)\mathds{1}_{\Omega_d}(\vec{r}) \\
        & \tens{\mu_{r,a}}(\vec{r})                 && = \tens{\mu_r}(\vec{r}),
    \end{alignedat}
\end{equation}
noting $\mathds{1}_D$ the characteristic function of a set $D$. This annex problem can be easily solved making use of the Fresnel coefficients in the appropriate orthonormal basis $(\hat{\vec{p}}, \hat{\vec{s}},\vec{k}^+_{\downarrow}/k^+)$ (see Fig.~\ref{fig::grating_2D_model}), which basically extends the notion of TE/TM polarization. More explicitly, the Fresnel coefficients are
\begin{equation}\label{eq::coeff_Fresnel}
    \begin{alignedat}{3}
        & r_s && = \frac{\beta^+ - \beta^-}{\beta^+ + \beta^-},
        \quad\quad\quad
        && t_s = \frac{2\beta^+}{\beta^+ + \beta^-}, \\
        & r_p && = \frac{\beta^+\varepsilon_r^- - \beta^-\varepsilon_r^+}{\beta^+\varepsilon_r^- + \beta^-\varepsilon_r^+},
        \quad\quad\quad
        && t_p = \frac{2\beta^+\varepsilon_r^-}{\beta^+\varepsilon_r^- + \beta^-\varepsilon_r^+} \cdot
    \end{alignedat}
\end{equation}
Hence, denoting $Z^+ = \sqrt{\dfrac{\mu_0\mu_r^+}{\varepsilon_0\varepsilon_r^+}}$ the impedance of the superstrate, the fully $\hat{\vec{s}}$-polarized electric and $\hat{\vec{s}}$-polarized magnetic fields write resp.
\begin{equation}\label{eq::Es_Hs}
    \setlength{\arraycolsep}{1pt} 
    \left\{
    \begin{array}{llc}
        \vec{E}^{\text{inc}}_{\hat{\vec{s}}} & =     & \exp(i\vec{k}^+_{\downarrow} \cdot \vec{x})\,\hat{\vec{s}} \\[1mm]
        \vec{E}^r_{\hat{\vec{s}}}            & = r_s & \exp(i\vec{k}^+_{\uparrow}   \cdot \vec{x})\,\hat{\vec{s}} \\[1mm]
        \vec{E}^t_{\hat{\vec{s}}}            & = t_s & \exp(i\vec{k}^-_{\downarrow} \cdot \vec{x})\,\hat{\vec{s}}
    \end{array}
    \right.
    \quad \text{and} \quad
    \left\{
    \begin{array}{llc}
        \vec{H}^{\text{inc}}_{\hat{\vec{s}}} & = 1/Z^+   & \exp(i\vec{k}^+_{\downarrow} \cdot \vec{x})\,\hat{\vec{s}} \\[1mm]
        \vec{H}^r_{\hat{\vec{s}}}            & = r_p/Z^+ & \exp(i\vec{k}^+_{\uparrow}   \cdot \vec{x})\,\hat{\vec{s}} \\[1mm]
        \vec{H}^t_{\hat{\vec{s}}}            & = t_p/Z^+ & \exp(i\vec{k}^-_{\downarrow} \cdot \vec{x})\,\hat{\vec{s}}
    \end{array}
    \right. .
\end{equation}
Then the fully $\hat{\vec{p}}$-polarized electric field can be deduced:
\begin{equation}\label{eq::Ep}
    \setlength{\arraycolsep}{1pt}
    \left\{
    \begin{array}{llc}
        \vec{E}^{\text{inc}}_{\hat{\vec{p}}} & = -\vec{k}^+_{\downarrow} \times \vec{H}^i_{\hat{\vec{s}}} /(\omega\varepsilon_0\varepsilon_r^+) \\[1mm]
        \vec{E}^r_{\hat{\vec{p}}}            & = -\vec{k}^+_{\uparrow}   \times \vec{H}^r_{\hat{\vec{s}}} /(\omega\varepsilon_0\varepsilon_r^+) \\[1mm]
        \vec{E}^t_{\hat{\vec{p}}}            & = -\vec{k}^-_{\downarrow} \times \vec{H}^t_{\hat{\vec{s}}} /(\omega\varepsilon_0\varepsilon_r^-)
    \end{array}
    \right. .
\end{equation}
Finally the linearly polarized annex electric field $\vec{E}_a$ solution of the annex problem writes:
\begin{equation}\label{eq::annex_electric}
    \vec{E}_a = A_e(\cos\psi_i \vec{E}_{a,\hat{\vec{p}}} - \sin\psi_i \vec{E}_{a,\hat{\vec{s}}}),
\end{equation}
where
\begin{equation}\label{eq::annex_electric_polar}
    \vec{E}_{a,\{\hat{\vec{s}}, \hat{\vec{p}}\}} = (\vec{E}^{\text{inc}}_{\{\hat{\vec{s}}, \hat{\vec{p}}\}} + \vec{E}^r_{\{\hat{\vec{s}}, \hat{\vec{p}}\}}) \mathds{1}_{\mathcal{S}^+} + \vec{E}^t_{\{\hat{\vec{s}}, \hat{\vec{p}}\}} \mathds{1}_{\mathcal{S}^-}.
\end{equation}
This last expression completes the tools needed to compute the response of a periodic structure lying upon a substrate through the scattered field formulation. It allows to write the solution of the Helmholtz equation \eqref{eq::total_field_PDE} as an outgoing field with a source localized into the design region. This outgoing field condition is modelled numerically by the PMLs that absorb the field radiating from the design region. Homogeneous Neumann or Dirichlet conditions on the lower and upper boundaries of $\Omega$ can be chosen to truncate the PMLs. The former allows to keep track of the field values at the PML's endings, while the latter (chosen here) sets the field to zero at the PML's endings while reducing the number of FEM unknowns. The corresponding functional space becomes $\mathcal{V} \coloneqq \mathcal{H}_0(\rot_{\gamma}, \Omega, e^{i\alpha d})$ in order to point out these extra boundary conditions. The scattered field formulation of the direct problem reads:
\begin{flushleft}
    \fcolorbox{white}{bleuclair}{
    \begin{minipage}[t][2.5cm][c]{.965\textwidth}
        \begin{proposition}
            The diffracted field $\vec{E}_*^d \coloneqq \vec{E}^{\text{tot}} - \vec{E}_a$ is the solution of the following PDE:
            \medbreak
            \noindent Find $\vec{E}_*^d \in \boldsymbol{\mathcal{V}}$ such that $\forall\vec{r}\in\Omega$
            with $\omega\in\mathbb{R}^+$ given,
            \begin{equation}\label{eq::scattered_field_PDE}
                \rot_{\gamma}(\tens{\mu_r}^{-1}\rot_{\gamma}\vec{E}_*^d) - k_0^2\tens{\varepsilon_r}\vec{E}_*^d
                = k_0^2 (\tens{\varepsilon_r} - \tens{\varepsilon_{r,a}})\vec{E}_a.
            \end{equation}
        \end{proposition}
    \end{minipage}
    }
\end{flushleft}
\medbreak
Since by construction $\tens{\varepsilon_r} - \tens{\varepsilon_{r,a}} = 0$ out of $\Omega_d$, the source (the so-called Right-Hand Side, RHS) is localized in the design region, which indeed guarantees the outgoing nature of the chosen diffracted field.

\subsection{Weak formulation}

The decomposition into a tangential and longitudinal part also applies to the characteristics of the materials. The permittivity and permeability tensors are decoupled in the following manner:
\begin{equation}\label{eq::permittivities_decomposition}
    \tens{\varepsilon_r} =
    \underbrace
    {\begin{pmatrix}
        \tilde{\tens{\varepsilon_r}} & 0 \\
        0                            & 1
    \end{pmatrix}}_{\text{\normalsize $\tens{\varepsilon_{r,t}}$}}
    \underbrace
    {\begin{pmatrix}
        \tens{I_2} & 0 \\
        0          & \varepsilon_{r,zz}
    \end{pmatrix}}_{\text{\normalsize $\tens{\varepsilon_{r,\ell}}$}} \quad \text{and} \quad
    \tens{\mu_r} =
    \underbrace
    {\begin{pmatrix}
        \tilde{\tens{\mu_r}} & 0 \\
        0                    & 1
    \end{pmatrix}}_{\text{\normalsize $\tens{\mu_{r,t}}$}}
    \underbrace
    {\begin{pmatrix}
        \tens{I_2} & 0 \\
        0          & \mu_{r,zz}
    \end{pmatrix}}_{\text{\normalsize $\tens{\mu_{r,\ell}}$}}
\end{equation}
so that $\tens{\varepsilon}\vec{E} = \tens{\varepsilon_t}\vec{E}_t + \tens{\varepsilon_{\ell}}E_{\ell} \hat{\vec{z}}$. Gathering all the definitions above and using the identity $\rot_{\gamma} \vec{E} = \rot_t \vec{E}_t + \left(\grad_t E_{\ell} - i\gamma \vec{E}_t\right) \times \hat{\vec{z}}$, the following weak formulation in the conical case is obtained:
\begin{flushleft}
    \fcolorbox{white}{bleuclair}{
    \begin{minipage}[t][6.2cm][c]{.965\textwidth}
        \begin{corollary}\label{cor::direct_problem}
            The weak formulation of the direct problem in the conical case is:
            \medbreak
            \noindent Find $\vec{E}^d_* \in \boldsymbol{\mathcal{V}}$ such that for all
            $\vec{E}' \in \boldsymbol{\mathcal{V}}$,
            \begin{equation}\label{eq::weak_formulation}
                \begin{alignedat}{1}
                    \int_{\Omega}
                    \Biggl[
                    \mu_{r,zz}^{-1} & \rot_t\vec{E}_{*,t}^d \cdot \overline{\rot_t\vec{E}_t'}
                    + \left(\tens{\mu_{r,t}}^{-1} (\grad_t E_{*,\ell}^d \times \hat{\vec{z}}) \right) \cdot \overline{\grad_t E_{\ell}' \times \hat{\vec{z}}} \\
                    & + i\gamma \left(\tens{\mu_{r,t}}^{-1} (\grad_t E_{*,\ell}^d \times \hat{\vec{z}}) \right) \cdot \overline{\vec{E}_t' \times \hat{\vec{z}}}
                    - i\gamma \left(\tens{\mu_{r,t}}^{-1} (\vec{E}_{*,t}^d \times \hat{\vec{z}}) \right) \cdot \overline{\grad_t E_{\ell}' \times \hat{\vec{z}}} \\
                    & + \gamma^2 \left(\tens{\mu_{r,t}}^{-1} (\vec{E}_{*,t}^d \times \hat{\vec{z}}) \right) \cdot \overline{\vec{E}_t' \times \hat{\vec{z}}}
                    - k_0^2 \left(\tens{\varepsilon_{r,t}}\vec{E}_{*,t}^d \cdot \overline{\vec{E}_t'} + \varepsilon_{r,zz} E_{*,\ell}^d \overline{E_{\ell}'}\right) \\
                    & + \underbrace{k_0^2 \left((\tens{\varepsilon_{r,a,t}} - \tens{\varepsilon_{r,t}})\vec{E}_{a,t} \cdot \overline{\vec{E}_t'} + (\varepsilon_{r,a,zz} - \varepsilon_{r,zz})E_{a,\ell} \overline{E_{\ell}'}\right)}_{\text{\emph{RHS direct problem}}}
                    \Biggr]
                    \, \mathrm{d}\Omega
                    = 0.
                \end{alignedat}
            \end{equation}
        \end{corollary}
    \end{minipage}
    }
\end{flushleft}
\medbreak
The decomposition of the periodic part of the solution $\vec{E}_*^d e^{-i\alpha x}$ into Fourier series allows to obtain the complex amplitudes of each diffraction order
\cite{book_fresnel_c5}:
\begin{equation}\label{eq::complex_amplitudes}
    \left\{
    \begin{array}{llc}
        \displaystyle r_n^u = \dfrac{1}{d}\int_0^d e^{-i\alpha_n x}\,\vec{E}^d(x,y_0)\cdot\hat{\vec{u}}\,\mathrm{d}x & \mbox{for } y_0 > y_{\Omega_d} & \mbox{(above the design region)} \\
        \displaystyle t_n^u = \dfrac{1}{d}\int_0^d e^{-i\alpha_n x}\,\vec{E}^{\text{tot}}(x,y_0)\cdot\hat{\vec{u}}\,\mathrm{d}x & \mbox{for } y_0 < 0 & \mbox{(below the design region)}
    \end{array}
    \right. ,
\end{equation}
where $u$ designates $x$, $y$ or $z$ and $\vec{E}^d = \vec{E}_*^d + \vec{E}_a^d$ is the total diffracted field. As a reminder, $\vec{E}^{\text{tot}} = \vec{E}_*^d + \vec{E}_a$. In practice, these complex amplitudes are evaluated resp. on the PML/superstrate and on the PML/substrate interfaces. Making use of the Poynting theorem, these amplitudes lead to the reflection and transmission efficiencies
\begin{equation}\label{eq::efficiencies}
    \left\{
    \begin{array}{llc}
        R_n = \dfrac{\beta_n^+}{A_e^2\beta^+}(\lvert r_n^x\rvert^2 + \lvert r_n^y\rvert^2 + \lvert r_n^z\rvert^2) \\[3mm]
        T_n = \dfrac{\beta_n^-}{A_e^2\beta^+}(\lvert t_n^x\rvert^2 + \lvert t_n^y\rvert^2 + \lvert t_n^z\rvert^2)
    \end{array}
    \right. .
\end{equation}
As the transverse component $E^d_{*,y} \coloneqq \vec{E}^d_{*,t} \cdot \hat{\vec{y}}$ is discontinuous across the superstrate/PML (resp. substrate/PML) interface, the quantity $r_n^y$ (resp. $t_n^y$) cannot be simply postprocessed from $\vec{E}^d\cdot\hat{\vec{y}}$ (resp. $\vec{E}^{\text{tot}}\cdot\hat{\vec{y}}$). It is still possible to access it by using a Lagrange multiplier mapping the normal trace of the transverse field $\vec{E}^d_{*,t}$ on the PML/superstrate (resp. PML/substrate) boundary. Nonetheless, considering the adjoint problem in section~\ref{sec::adjoint_method}, it is more appropriate to use another expression of the diffraction efficiencies that only involves the components of the field tangential to the superstrate/PML (resp. substrate/PML) interface \cite{theseZolla}:
\begin{equation}\label{eq::alternate_Rn}
    \left\{
        \begin{array}{llc}
            R_n = \dfrac{1}{\beta_n^+\beta^+A_e^2}&\Bigl[((\beta_n^+)^2 + \alpha_n^2)\lvert r_n^x\rvert^2
                                            + ((\beta_n^+)^2 + \gamma^2)\lvert r_n^z\rvert^2
                                            + 2\alpha_n\gamma\text{Re}(r_n^x\overline{r_n^z})
                                        \Bigr] \\[2mm]
            T_n = \dfrac{1}{\beta_n^-\beta^+A_e^2}&\Bigl[((\beta_n^-)^2 + \alpha_n^2)\lvert t_n^x\rvert^2
                                            + ((\beta_n^-)^2 + \gamma^2)\lvert t_n^z\rvert^2
                                            + 2\alpha_n\gamma\text{Re}(t_n^x\overline{t_n^z})
                                        \Bigr]
        \end{array}
    \right. .
\end{equation}
From now on, the axes $x$ and $z$ are gathered with the notation $u$. For example, writing $r_n^u$ stands for $r_n^x$ \emph{and} $r_n^z$.

\section{Optimization problem}\label{sec::optimization_problem}

\subsection{Design variables and constraints}

Let us consider a density (or design) variable $\boldsymbol{\rho}$ constant per element of the mesh. Let thus assume that if the mesh is made of $N$ triangles, $\boldsymbol{\rho}$ is a vector of size $N$ and that $0 \leq \rho_i \leq 1$ for all $i$. Since this density field defines the relative permittivity of the design space, changing $\boldsymbol{\rho}$ modifies the structure. It thus has a direct impact on the scattered field $\vec{E}_*^d$ and the diffraction efficiency $R_n$. From now on, the notation $R_n$ designates a function $R_n: [0,1]^N \ni \boldsymbol{\rho} \mapsto R_n(\boldsymbol{\rho}) \in [0,1]$.

The \emph{connectedness constraint} is imposed using a connectedness per element map $\boldsymbol{\rho}_f$, which is a sliding averaging \cite{TD_3D_TopOpt} as detailed in the supplementary of this article. The \emph{binarization} is applied over this first filter and obtained using the usual function \cite{ADCode} called $\boldsymbol{\hat{\rho}}$:
\begin{equation}\label{eq::filter}
    \boldsymbol{\hat{\rho}}: [0,1]^N \ni \boldsymbol{\rho}_f
    \mapsto \frac{\tanh(\beta_f\nu) + \tanh[\beta_f(\boldsymbol{\rho}_f - \nu)]}
           {\tanh(\beta_f\nu) + \tanh[\beta_f(1 - \nu)]} \in [0,1]^N
\end{equation}
with $\nu\in[0,1]$ and where $\beta_f$ is increased during the optimization process. A standard configuration is $\nu = 1/2$ and $\beta_{f,m} = 2^m$ with $m = 1,\dots,7$ increasing during the optimization process.

To summarize, the constraints on the design variables are gathered in one function that coerces the design into having a specific behaviour. It can be written through a composition of maps:
\begin{equation}\label{eq::comp_epsilon}
    \boldsymbol{\hat{\rho}}_f: [0,1]^N \ni \boldsymbol{\rho} \mapsto (\boldsymbol{\hat{\rho}} \circ \boldsymbol{\rho}_f)(\boldsymbol{\rho}) \in [0,1]^N.
\end{equation}
Now we are in position to introduce the optimization problem.

\subsection{Optimization of diffraction efficiencies}

The objective of the study is to provide a blazed metasurface which is the most efficient as possible on one particular diffraction order in reflection. Note that extending what follows to the case of transmission is straightforward. Therefore we want to maximize the energy related quantity $R_n$ of Eq.~\eqref{eq::alternate_Rn} for a specific value of $n$ on the constrained density field $\boldsymbol{\hat{\rho}}_f$. Let then define the composed function $\mathcal{R}_n: [0,1]^N \ni \boldsymbol{\rho} \mapsto (R_n \circ \boldsymbol{\hat{\rho}} \circ \boldsymbol{\rho}_f)(\boldsymbol{\rho}) \in [0,1]$ that returns the diffraction efficiency in reflection induced by the constrained density distribution. We choose to minimize $\mathcal{F}_n \coloneqq 1 - \mathcal{R}_n$ since $\mathcal{R}_n$ has values between 0 and 1. The DoFs are the densities $\rho_i$ into the design region with $i\in\llbracket 1,N\rrbracket$. The optimization problem writes:
\begin{equation}\label{eq::opt_prob}
    \begin{alignedat}{2}
        \underset{\boldsymbol{\rho}}{\min} &                && \mathcal{F}_n(\boldsymbol{\rho}) \\
        \text{such that}                   & \text{ \ \ \ } && 
        \left\{ \begin{array}{llc}
                    \mathcal{L}(\boldsymbol{\rho}, \vec{E}') = 0 & \mbox{$\forall \vec{E}' \in \boldsymbol{\mathcal{V}}$} \\
                    0 \leq \rho_i \leq 1 & \mbox{$\forall i\in\llbracket 1,N \rrbracket$}
                \end{array}
        \right. ,
    \end{alignedat}
\end{equation}
where $\mathcal{L}$ designates the weak formulation in Eq.~\eqref{eq::weak_formulation}, that is, written without the tangential/longitudinal decomposition for compactness:
\begin{equation}\label{eq::constr_L}
    \begin{alignedat}{1}
        \mathcal{L} :  \mathbb{R}^N \times \boldsymbol{\mathcal{V}} \ni (\boldsymbol{\rho},\vec{E}')
        \mapsto 
        \int_{\Omega} &\bigg[
                    \tens{\mu_r}^{-1} \rot_{\gamma}\vec{E}_*^d(\boldsymbol{\rho}) \cdot \rot_{\gamma}\overline{\vec{E}'} \\
                    &- k_0^2\Big(\tens{\varepsilon_r}(\boldsymbol{\rho})\vec{E}_*^d(\boldsymbol{\rho}) + (\tens{\varepsilon_r}(\boldsymbol{\rho}) - \tens{\varepsilon_{r,a}})\vec{E}_a\Big) \cdot \overline{\vec{E}'}
        \bigg] \mathrm{d}\Omega \in \mathbb{R}.
    \end{alignedat}
\end{equation}
Note that $\vec{E}_*^d$ is the solution of Eq.~\eqref{eq::scattered_field_PDE} for a given design variable $\boldsymbol{\rho}_*$. In other terms, $\forall\vec{E}'\in\boldsymbol{\mathcal{V}},\,\mathcal{L}(\boldsymbol{\rho}_*, \vec{E}') = 0$. This is why $\boldsymbol{\rho}_*$ is called the equilibrium point in the framework of optimization. Given the number of unknowns, this kind of optimization problem is intractable without the Jacobian of the target $\mathcal{F}_n$ with respect to the design variable $\boldsymbol{\rho}$.

\subsection{Jacobian of the target}

As mentioned above, the target function is actually a composition of three maps, namely:
\begin{equation}\label{eq::comp_target}
    \mathcal{F}_n(\boldsymbol{\rho}) = (F_n \circ \boldsymbol{\hat{\rho}} \circ \boldsymbol{\rho}_f)(\boldsymbol{\rho}),
\end{equation}
where $F_n = 1 - R_n$. The chain rule is applied to get the expression of the derivatives with respect to all $\rho_i$:
\begin{equation}\label{eq::target_derivative_general}
    \frac{\partial\mathcal{F}_n}{\partial\boldsymbol{\rho}}(\boldsymbol{\rho})
    = \frac{\partial\boldsymbol{\rho}_f}{\partial\boldsymbol{\rho}}(\boldsymbol{\rho})
    \frac{\partial\boldsymbol{\hat{\rho}}}{\partial\boldsymbol{\rho}_f}(\boldsymbol{\rho}_f)
    \frac{\partial F_n}{\partial\boldsymbol{\hat{\rho}}_f}(\boldsymbol{\hat{\rho}}_f).
\end{equation}
The notation (bold or not) shows whether the Jacobian is a vector of size $N$ (differentiation of a scalar with respect to a vector) or an $N\times N$ matrix (differentiation of a vector with respect to a vector). The first two factors are known since the definitions of the constraints on the design variable are differentiable analytic functions (see the technical support). Moreover starting from Eq.~\eqref{eq::alternate_Rn} and recalling that $\partial \lvert r_n^u \rvert^2 / \partial\hat{\rho}_i = 2\mathrm{Re}\{\overline{r_n^u} \, \partial r_n^u / \partial \hat{\rho}_i\}$ \cite{bao_inverseprob}, we obtain:
\begin{equation}\label{eq::derivative_Rn}
    \frac{\partial F_n}{\partial \hat{\rho}_{f,i}}
    = - \frac{\partial R_n}{\partial \hat{\rho}_{f,i}} = -\frac{2}{\beta_n^+\beta^+A_e^2}
    \text{Re}\Biggl[((\beta_n^+)^2 + \alpha_n^2)\frac{\partial r_n^x}{\partial \hat{\rho}_{f,i}}           \overline{r_n^x}
            + ((\beta_n^+)^2 + \gamma^2)\frac{\partial r_n^z}{\partial \hat{\rho}_{f,i}}\overline{r_n^z}
            + \alpha_n\gamma\Bigl(\frac{\partial r_n^x}{\partial \hat{\rho}_{f,i}}\overline{r_n^z} + r_n^x\frac{\partial \overline{r_n^z}}{\partial \hat{\rho}_{f,i}}\Bigr)
        \Biggr].
\end{equation}
However, the complex amplitudes $r_n^u$, $u = \{x,z\}$, depend explicitly on the diffracted field $\vec{E}_*^d$ that is only computed numerically using the FEM. Consequently the analytic derivatives are not available. An intuitive but extremely costly solution would consist in computing the derivative numerically using finite differences for each design variable $\rho_i$. This would require solving the direct problem $N$ times per iteration of the optimization process, which is clearly prohibitive. This issue led to the development of the so-called adjoint method.

\section{Adjoint method}\label{sec::adjoint_method}

The last hurdle to compute numerically the Jacobian in Eq.~\eqref{eq::derivative_Rn} is to know the value of $\partial r_n^u / \partial \hat{\rho}_{f,i}$ around the current equilibrium point $\boldsymbol{\rho}_*$. It is provided by the following proposition:
\begin{flushleft}
    \fcolorbox{white}{bleuclair}{
    \begin{minipage}[t][6.9cm][c]{.965\textwidth}
        \begin{proposition}\label{prop::adjoint}
            The derivatives of $r_n^u$, $u = \{x,z\}$, with respect to the $\hat{\rho}_{f,i}$ around the
            equilibrium point $\boldsymbol{\hat{\rho}}_{f,*}$ are given by
            \begin{equation}\label{eq::adjoint_derivative_rn}
                \frac{\partial r_n^u}{\partial \hat{\rho}_{f,i}} (\boldsymbol{\hat{\rho}}_{f,*})
                = \int_{\mathcal{T}_i} k_0^2 (\varepsilon_{r,\text{diel}} - \varepsilon_r^+) \vec{E}^{\text{tot}} (\boldsymbol{\rho}_*) \cdot \boldsymbol{\lambda}_*^u \, \mathrm{d}\Omega,
            \end{equation}
            where $\mathcal{T}_i$ is the triangle of the mesh with the
            density $\hat{\rho}_{f,i}$ and $\boldsymbol{\lambda}_*^u$ is the unique solution of the \emph{adjoint problem}:
            \medbreak
            \noindent Find
            $\boldsymbol{\lambda}_*^u\in\boldsymbol{\mathcal{V}}^{\text{adj}}= \boldsymbol{\mathcal{H}}_0(\rot_{-\gamma}, \Omega, e^{-i\alpha d})$
            such that for all $\boldsymbol{\lambda}' \in
            \boldsymbol{\mathcal{V}}^{\text{adj}}$,
            \begin{equation}\label{eq::adjoint_problem}
                \int_{\Omega}
                \Biggl[
                \tens{\mu_r}^{-1} \rot_{\gamma}\boldsymbol{\lambda}_*^u \cdot \rot_{\gamma}\overline{\boldsymbol{\lambda}'}
                - k_0^2 \tens{\varepsilon_r}\boldsymbol{\lambda}_*^u \cdot \overline{\boldsymbol{\lambda}'}
                \Biggr]\,\mathrm{d}\Omega
                = \frac{1}{d} \int_0^d e^{-i\alpha_n x} \hat{\vec{u}} \cdot\overline{\boldsymbol{\lambda}'} \,\mathrm{d}x.
            \end{equation}
            The RHS is defined on a line [0,d] in the superstrate.
        \end{proposition}
    \end{minipage}
    }
\end{flushleft}
\medbreak
The proof of this proposition is detailed in the technical support of this document. Note that if the domain $\Omega$ is 3D, then the adjoint problem Eq.~\eqref{eq::adjoint_problem} remains the same, but the source becomes an integral over a surface $\Gamma^+$ in the superstrate. The following corollary provides the weak formulation in order to solve Eq.~\eqref{eq::adjoint_problem} with the FEM in conical mounting.
\begin{flushleft}
    \fcolorbox{white}{bleuclair}{
    \begin{minipage}[t][6.3cm][c]{.965\textwidth}
        \begin{corollary}\label{cor::adjoint_problem}
            The weak formulations of the adjoint problems on $u = \{x,z\}$ in the conical case are:
            \medbreak
            \noindent Find $\boldsymbol{\lambda}_*^u \in
            \boldsymbol{\mathcal{V}}^{\text{adj}}$ such that for all $\boldsymbol{\lambda}'
            \in \boldsymbol{\mathcal{V}}^{\text{adj}}$,
            \begin{equation}\label{eq::weak_formulation_adj}
                \begin{alignedat}{1}
                    \int_{\Omega}
                    \Biggl[
                    \mu_{r,zz}^{-1} & \rot_t\boldsymbol{\lambda}_{*,t}^u \cdot \overline{\rot_t\boldsymbol{\lambda}_t'}
                    + \left(\tens{\mu_{r,t}}^{-1} (\grad_t \lambda_{*,\ell}^u \times \hat{\vec{z}}) \right) \cdot \overline{\grad_t \lambda_{\ell}' \times \hat{\vec{z}}} \\
                    & - i\gamma \left(\tens{\mu_{r,t}}^{-1} (\grad_t \lambda_{*,\ell}^u \times \hat{\vec{z}}) \right) \cdot \overline{\boldsymbol{\lambda}_t' \times \hat{\vec{z}}}
                    + i\gamma \left(\tens{\mu_{r,t}}^{-1} (\boldsymbol{\lambda}_{*,t}^u \times \hat{\vec{z}}) \right) \cdot \overline{\grad_t \lambda_{\ell}' \times \hat{\vec{z}}} \\
                    & + \gamma^2 \left(\tens{\mu_{r,t}}^{-1} (\boldsymbol{\lambda}_{*,t}^u \times \hat{\vec{z}}) \right) \cdot \overline{\boldsymbol{\lambda}_t' \times \hat{\vec{z}}}
                    - k_0^2 \left(\tens{\varepsilon_{r,t}}\boldsymbol{\lambda}_{*,t}^d \cdot \overline{\boldsymbol{\lambda}_t'} + \varepsilon_{r,zz} \lambda_{*,\ell}^d \overline{\lambda_{\ell}'}\right) \\
                    & - \underbrace{\frac{1}{d} \int_0^d e^{-i\alpha_n x}  \Bigl( \overline{\boldsymbol{\lambda}_t'} + \overline{\lambda_{\ell}'}\hat{\vec{z}} \Bigr) \cdot \hat{\vec{u}} \,\mathrm{d}x}_{\text{\emph{RHS adjoint problem}}}
                    \Biggr] \, \mathrm{d}\Omega = 0.
                \end{alignedat}
            \end{equation}
        \end{corollary}
    \end{minipage}
    }
\end{flushleft}
\medbreak
Note that the only differences with Eq.~\eqref{eq::weak_formulation} of the Corollary~\ref{cor::direct_problem} are the signs before the $i\gamma$ terms  on the one hand, and the RHS on the other hand. This is justified in the technical support.

With the Corollary~\ref{cor::direct_problem}, the direct problem is solved thanks to its weak formulation Eq.~\eqref{eq::weak_formulation} in order to compute $\vec{E}_*^d$ and thus $\vec{E}^{\text{tot}}$. With the Corollary~\ref{cor::adjoint_problem}, the adjoint problems on both the $x$ and $z$ axes are solved with the weak formulation Eq.~\eqref{eq::weak_formulation_adj} in order to compute $\boldsymbol{\lambda}_*^u$, $u=\{x,z\}$. Therefore, the derivatives of $r_n^u$ are known using Eq.~\eqref{eq::adjoint_derivative_rn}. It eventually enables to reach the derivatives of the target function $\mathcal{F}_n$ through Eq.~\eqref{eq::derivative_Rn} and Eq.~\eqref{eq::target_derivative_general}.

At first sight, the calculation of the derivatives of $\mathcal{F}_n$ necessitates three Finite Element resolutions per iteration of the optimization process: the direct problem and two adjoint problems. However, the Finite Element matrix for both adjoint problems is strictly identical, only the RHS is changed. Therefore, as soon as this matrix is constructed and inverted (which represents the most costly part of a FEM run) to solve the adjoint problem on $x$, it can be properly re-used for the adjoint problem on $z$. Consequently, only two costly matrix inversions are needed per iteration of the optimization process, one for the direct and one for both adjoint problems.
\newpage
\section{Discretization and numerical aspects}\label{sec::discretization_aspects}

The following discrete spaces are used in the simulations:
\begin{itemize}
    \item The scattered vector field $\vec{E}^d_*\in\boldsymbol{\mathcal{V}}$ is by construction split into its possibly discontinuous transverse components and longitudinal continuous one: $\vec{E}^d_*=\vec{E}_{*,t}^d+E_{*,\ell}^d\,\hat{\vec{z}}$. The transverse vector field $\vec{E}_{*,t}^d$ is discretized using hierarchical Webb elements \cite{geuzaine1999convergence,webb1993hierarchal,jin02FEM-electromag}, with 3 FEM unknowns per edge and 2 per face, that is 11 FEM unknowns per triangle. The longitudinal scalar field $E_{*,\ell}^d$ is discretized using Lagrange elements of the second order (denoted $P^2$), that is 6 FEM unknowns per triangle. Bloch boundary conditions are applied to both discrete spaces: the unknowns defined on the right boundary of $\Omega$ ($x=d$) are the same as those defined on the left boundary ($x=0$) up to a phase shift $e^{+i\alpha d}$. Dirichlet boundary conditions are applied to the top and bottom boundaries of resp. $\text{PML}^+$ and $\text{PML}^-$.
    \item The discrete version of the functional space $\boldsymbol{\mathcal{V}}^{\text{adj}}$ on which the adjoint vector field $\boldsymbol{\lambda}_*^u$ is defined follows the exact same steps but one: the phase shift for the Bloch boundary conditions is $e^{-i\alpha d}$.
    \item The densities and the Jacobian are constant scalars per mesh triangles. For the Jacobian, following the method described in Ref.~\cite{ShapeOpt_Geuzaine}, integrals over each mesh element defined in Eq.~\eqref{eq::adjoint_derivative_rn} are in fact performed by solving a weak projection of the integrand on the discontinuous constant per element space ($P^0$ elements). This projection corresponds to a trivial weak formulation defined on the design space, which is easy and fast to run.
\end{itemize}
The direct and adjoint problems are solved in parallel with the GetDP software \cite{getdp} on a mesh generated by Gmsh \cite{gmsh}. The optimization is led with the GCMMA \cite{GCMMA} using the NLopt package \cite{nlopt}. The accuracy of the numerical scheme is detailed in the technical support, with an even more general function: a multi-wavelength target, defined in the subsection~\ref{sec::multiwavelength} of this paper.

\section{Optimized blazed metasurfaces}\label{sec::examples}

\subsection{Patterning a silica slab above a silver substrate}

\subsubsection{Mono-wavelength optimization}

Here the study is focused on an application of the method on a concrete example, using silica (relative permittivity described in Ref.~\cite{fused_silica}) over a silver substrate (relative permittivity described in Ref.~\cite{silver_JC}). The period $d$ is set up to 3300\,nm and for now a single incident plane wave is considered, with wavelength $\lambda = $ 700\,nm and the incident angles $\theta_i = 5^{\circ}$, $\varphi_i = -66^{\circ}$ and $\psi_i = 90^{\circ}$. These parameters can be found in experiments such as in Ref.~\cite{BATMAN_instrument}.

\begin{figure}[ht!]
    \centering
    \includegraphics[width=.86\textwidth]{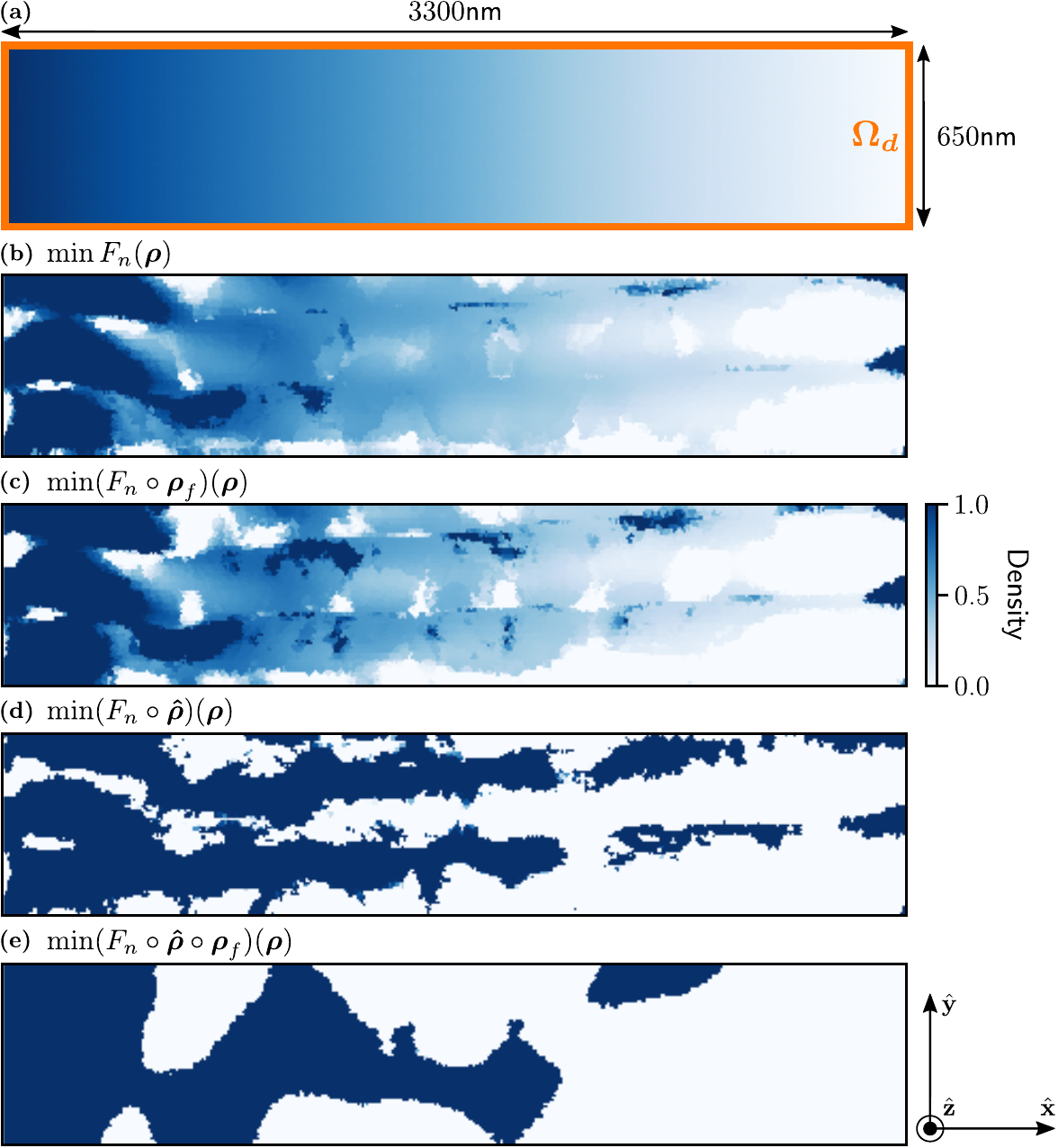}
    \caption{Optimization of the pattern on the $-1$ diffraction order for a period of 3300\,nm, a target wavelength of 700\,nm, the angles $\theta_i = 5^{\circ}$, $\varphi_i = -66^{\circ}$, $\psi_i = 90^{\circ}$, an oscillation tolerance of the target function of $10^{-5}$ and 15726 optimization DoFs in the design region. Four steps of constraints are displayed. \textbf{(a)}~Initial configuration chosen for this optimization: a graded-indexed blazed "silica to air" layer of dielectric with a thickness of 650\,nm and a linear spatial variation of the density from 1 (left) to zero (right); \textbf{(b)}~no constraint, $R_{-1} = 98.3\%$ reached in 18 iterations; \textbf{(c)}~only connectedness (filter radius $r_f = 200$\,nm), $R_{-1} = 98\%$ reached in 100 iterations; \textbf{(d)}~only binarization, $R_{-1} = 99\%$ reached in 503 iterations, and \textbf{(e)}~with binarization and connectedness, $R_{-1} = 97.9\%$ reached in 492 iterations.}
    \label{fig::silica_opt}
\end{figure}

\begin{figure}[ht!]
    \centering
    \includegraphics[width=1\textwidth]{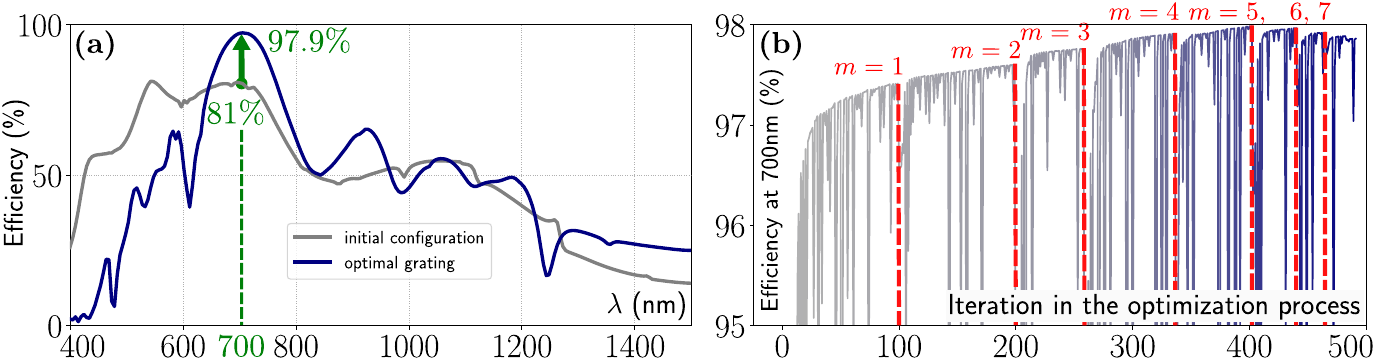}
    \caption{Optimization process leading to the pattern of Fig.~\ref{fig::silica_opt}\subFigRef{e}. \textbf{(a)}~Spectral response of the $-1^{\text{st}}$ diffraction efficiency in reflection on the [400,1500]\,nm spectral range under the target incidence angles. The maximum efficiency is reached exactly at the targeted wavelength 700\,nm, highlighted by the green dashed line. The green arrow illustrates the improvement enabled by the optimization algorithm on this very wavelength, starting from the initial configuration which spectral response is represented in grey; \textbf{(b)}~Evolution of the $-1^{\text{st}}$ diffraction efficiency at 700\,nm during the optimization process, zoom on the [95,98]\% efficiency interval. Each drop corresponds to an increment of the binarization process $m$ which are imposed at the red dashed lines.}
    \label{fig::efficiency_monowavelength}
\end{figure}

The initial configuration of the optimization process remains to be discussed. Indeed, given the large number of DoFs, the function $\mathcal{F}_n(\boldsymbol{\rho})$ exhibits many local minima, and the initial configuration has a large impact on the resulting local minimum found at the end of the iterative process. We choose to start from a non-realistic blazed grating designed to reproduce the same phase shift as the one induced by a blazed sawtooth silver grating with a blaze angle of $5^{\circ}$ and a period $d =$ 3300\,nm. This equivalent layer has a thickness of 650\,nm with a linear graded permittivity distribution as shown in Fig.~\ref{fig::silica_opt}\subFigRef{a} and detailed in the technical support. Such a configuration already provides an efficiency in the blazed order of 81\% at $\lambda = 700$\,nm, as shows its spectral response in grey color in Fig.~\ref{fig::efficiency_monowavelength}\subFigRef{a}.

The effect of each constraint is illustrated by running four optimizations at a single wavelength with the same initial configuration. The patterns obtained on one period as well as the number of iterations and the efficiency reached at the targeted wavelength $\lambda = 700$\,nm are shown in Fig.~\ref{fig::silica_opt}. First, neither the connectedness nor the binarization filter are used (Fig.~\ref{fig::silica_opt}\subFigRef{b}). Then the connectedness filter solely (Fig.~\ref{fig::silica_opt}\subFigRef{c}) or the binarization filter solely (Fig.~\ref{fig::silica_opt}\subFigRef{d}) is applied. Finally both constraints are taken into account (Fig.~\ref{fig::silica_opt}\subFigRef{e}). 

In Fig.~\ref{fig::silica_opt}, the level of blue designates the density of silica (between 0 and 1, 0 is white) on every triangle of the Finite Element mesh in one period of the design space. Without binarization, it leads to an unrealistic blurry (\emph{i.e.} graded-indexed) shape. The sole connectedness constraint slightly enlarges small details. The effect of the sole binarization is clear in Fig.~\ref{fig::silica_opt}\subFigRef{d} since a binary design is obtained. The combination of both constraints in Fig.~\ref{fig::silica_opt}\subFigRef{e} leads to a readable shape, although the manufacturability is not guaranteed due to the remaining free-standing substructure in the upper-right area of the cell, which is a known phenomenon given the connectedness filter used here \cite{topopt_tutorial}. Moreover, the application of the binarization filter above connectedness implies that these substructures may have a smaller size than $r_f$ \cite{TD_3D_TopOpt}. The optimum found is close to 100\% (99\%, given that 1\% of the total light is absorbed at 700\,nm by Joule effect). Therefore an important and encouraging remark is that neither the binarization nor the connectedness highly affects the maximal efficiency. Actually, another even higher efficiency maximum is found with the sole binarization. Indeed, the function to optimize and its Jacobian change for each constraint. It explains why there is no intuitive continuity between these results.

The spectral response on the $-1^{\text{st}}$ order as well as the evolution of the efficiency at the targeted wavelength of 700\,nm during the optimization process with binarization and connectedness are shown in Fig.~\ref{fig::efficiency_monowavelength}.

A maximum of efficiency is clearly reached (blue curve in Fig.~\ref{fig::efficiency_monowavelength}\subFigRef{a}) at the wavelength 700\,nm, which shows that the optimized result corresponds to a resonant mechanism at the targeted wavelength. The Fig.~\ref{fig::efficiency_monowavelength}\subFigRef{b} shows that the optimization is stable, and that the binarization process does perturb the optimization for a few iterations only (see the red dashed lines, showing each increment of the integer binarization parameter $m$ introduced in Eq.~\eqref{eq::filter}). Moreover, the threshold where the binarization prevents the efficiency to go higher is visible after $m=5$ (iteration 400), when the density $\rho$ can only take values really close to 0 or 1.

\begin{figure}[ht!]
    \centering
    \includegraphics[width=1\textwidth]{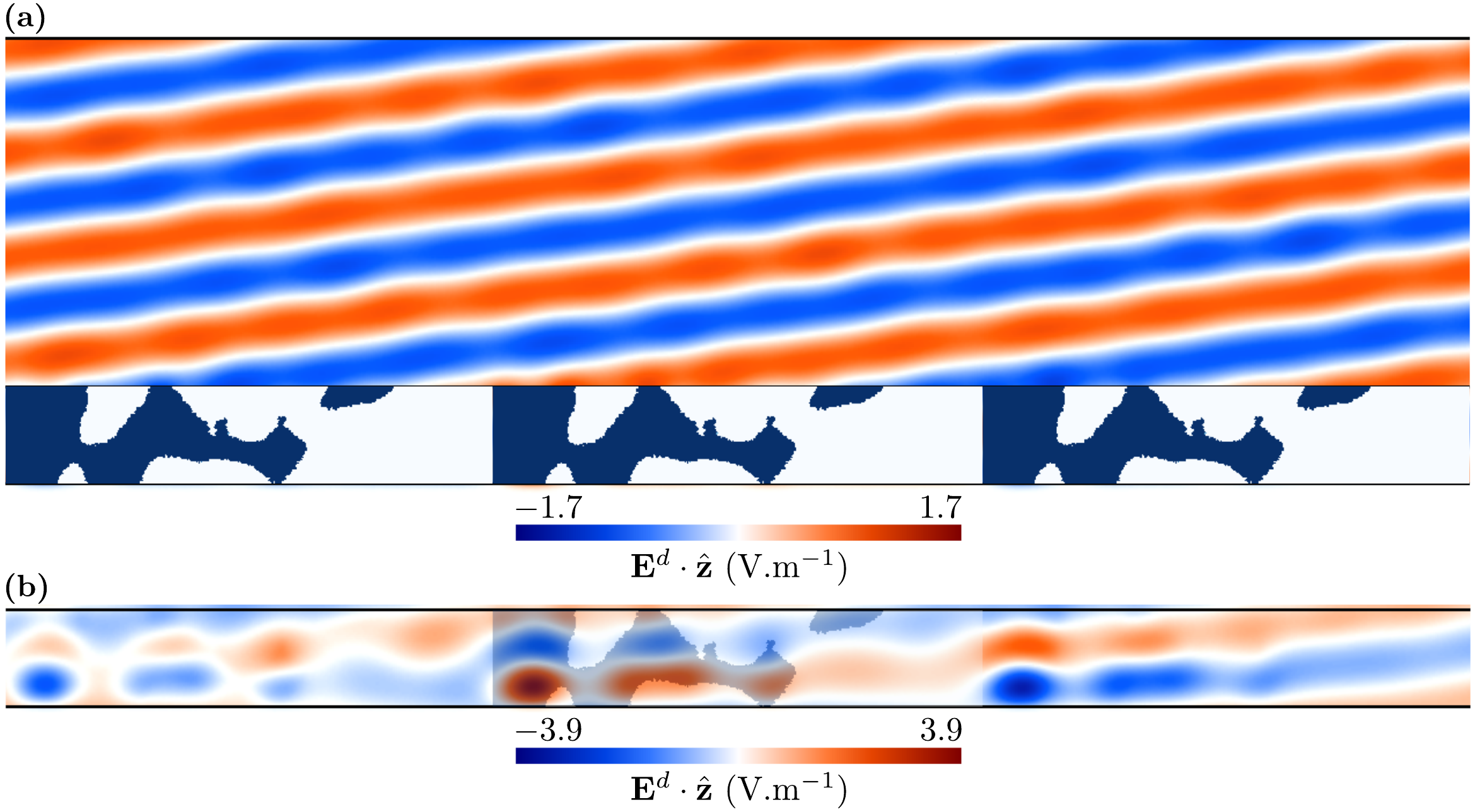}
    \caption{\textbf{(a)}~Diffracted field with the optimized pattern of Fig.~\ref{fig::silica_opt}\subFigRef{e} on three periods, for the target wavelength $\lambda = $ 700\,nm with the target incidence angles. It is close to a plane wave reflected with the angle $\theta_{-1} = -5^{\circ}$. \textbf{(b)}~Diffracted field in the groove region on the same three periods with the same specifications. Due to a strong resonance at the bottom left corner of each periodic cell, the scale of $\vec{E}^d\cdot\hat{\vec{z}}$ is not the same as in Fig.~\subFigRef{a}.}
    \label{fig::diffracted_field}
\end{figure}

The fact that the efficiency nearly reaches 100\% at this wavelength can be highlighted by displaying the corresponding diffracted field (projected on the $(Oxy)$ plane, \emph{i.e.} on the $z$ axis), as shown in Fig.~\ref{fig::diffracted_field}\subFigRef{a}. In the latter, three periods of the pattern and 3$\mu$m of the air above are visible. The target incidence angles are taken as well as the target wavelength. In particular, with $\theta_i = 5^{\circ}$, the angle of reflection on the~$-1^{\text{st}}$ diffraction order is $\theta_{-1} = -5^{\circ}$. The fact that the diffracted field is really close to a plane wave with a 5$^{\circ}$ deflection is a direct illustration of the 98\% efficiency of the pattern in this order. To complete this study, the field inside the design region is also displayed in Fig.~\ref{fig::diffracted_field}\subFigRef{b}, revealing a resonance at the bottom left corner of the design space.

With this mono-wavelength optimization process, a resonant topology was drilled into the design space. This resonant process has a quite low quality factor given the width of the resonance, but it is not broadband given the spectral range targeted. The reflection averaged over the spectral range [400,1500]\,nm (defined as $\int_{400}^{1500}R_{-1}(\lambda)\,\mathrm{d}\lambda/(1500-400)$) is only 51\%, whereas the sawtooth grating already reaches 52\%.

\subsubsection{Multi-wavelength optimization}\label{sec::multiwavelength}

\begin{figure}[ht!]
    \centering
    \includegraphics[width=1\textwidth]{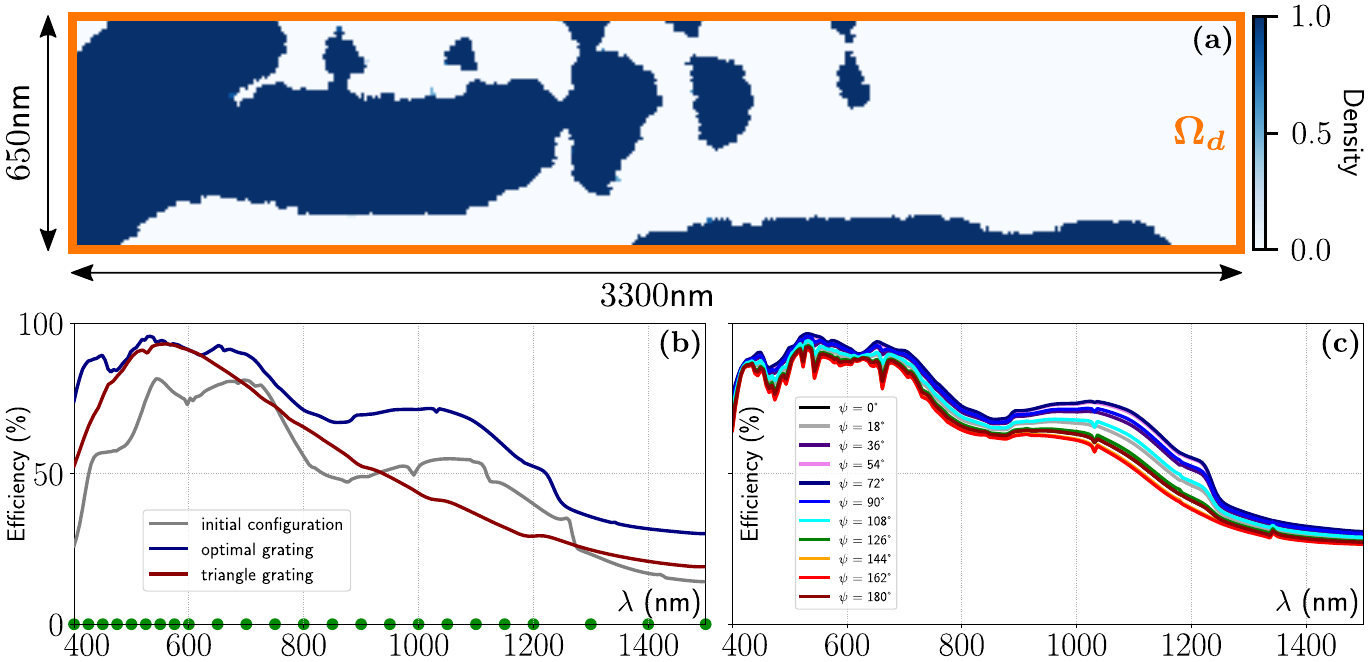}
    \caption{Results for the multi-wavelength optimization on the $-$1$^{\text{st}}$ diffraction order for a 650\,nm-high pattern on $N_{\lambda} = 24$ target wavelengths (green dots in Fig. b) between 400 and 1500\,nm, for the angles $\theta_i = 5^{\circ}$, $\varphi_i = -66^{\circ}$, $\psi_i = 90^{\circ}$, a solver tolerance of the target of $10^{-5}$, with connectedness (filter radius $r_f = 200$\,nm) and binarization. \textbf{(a)}~Optimized pattern. \textbf{(b)}~Spectral response in reflection on the [400,1500]\,nm spectral range under the target incidence angles. A comparison is displayed with the classical blazed grating (in red) and the initial configuration (in grey) with the same angles of incidence. \textbf{(c)} Spectral response in reflection with 11 evenly spaced polarization angles $\psi_i \in [0,180]^{\circ}$.}
    \label{fig::efficiency_multiwavelength}
\end{figure}

In order to broaden the spectral interval of the blaze effect, a multi-wavelength objective is now considered on $N_{\lambda}$ wavelengths:
\begin{equation}\label{eq::Fb_multilambda}
    F_n(\boldsymbol{\rho}) = 1 - \frac{1}{N_{\lambda}} \sum_{\lambda_i \in \Lambda} R_n(\boldsymbol{\rho}, \lambda_i).
\end{equation}
where $\Lambda = \{\lambda_1, \dots, \lambda_{N_{\lambda}}\}$ is a set of discrete targeted wavelengths chosen within the spectral range of interest. The density of targeted wavelengths is higher in spectral intervals exhibiting resonances (see green dots in abscissa of Fig.~\ref{fig::efficiency_multiwavelength}\subFigRef{b}).

Since each Jacobian of $R_n(\boldsymbol{\rho}, \lambda_i),\,\lambda_i\in\Lambda$ can be found following the steps described in the previous mono-wavelength case, the multi-wavelength Jacobian is trivial to compute. Note that one has to solve $N_\lambda$ direct problems along with $N_\lambda$ adjoint problems, which increases the computational burden. This is why the code has been parallelized to run all the wavelengths at once. As at this stage of the resolution, the problems are totally independent from a wavelength to another, there is no communication requirement for this parallelization.

The multi-wavelength optimization process on the initial configuration of Fig.~\ref{fig::silica_opt}\subFigRef{a} has a striking effect on the bandwidth of the blaze effect as shown in Fig.~\ref{fig::efficiency_multiwavelength}\subFigRef{b}. For this optimization, the target is composed of $N_{\lambda} = 24$ different target wavelengths, separated by 25\,nm on the interval [400,600]\,nm, by 50\,nm on the interval [600,1200]\,nm and 100\,nm on the interval [1200,1500]\,nm. More precisely, the diffraction efficiency averaged on the bandwidth is significantly increased, reaching 66\%, which is an absolute increasing of 14\% in comparison with the sawtooth grating (52\%) on the same wavelength interval.

In fact, the blaze response (blue curve in Fig.~\ref{fig::efficiency_multiwavelength}\subFigRef{b}) obtained is equal or higher than that of the sawtooth grating (red curve) over \emph{the whole} interval of interest. This higher performance result demonstrates the relevance of the multi-wavelength approach. 

Finally, we stress that although the multi-wavelength optimization was carried out for a particular incident polarization angle, the dependency of the response with respect to the polarization angle $\psi_i$ is moderate, as shown in Fig.~\ref{fig::efficiency_multiwavelength}\subFigRef{c} for multiple values of the polarization angle $\psi_i$ ranging in $[0,180]^{\circ}$. The maximal discrepancy between all the polarization angles reaches 17\% around 1050\,nm, while on the [400,1500]\,nm interval its averaged value is 8.4\%.

\subsection{Larger design space}\label{sec::largerdesign}

The longest wavelengths are not diffracted as efficiently as in the visible range, which is due to the thickness $y_{\Omega_d}$ of the design region. When $\lambda > y_{\Omega_d}$, the efficiencies keep dropping down because the material becomes sub-wavelength in the vertical direction, which is not enough to provide a sufficient phase shift. Therefore, the thickness of the design region is increased.

\begin{figure}[ht!]
    \centering
    \includegraphics[width=1\textwidth]{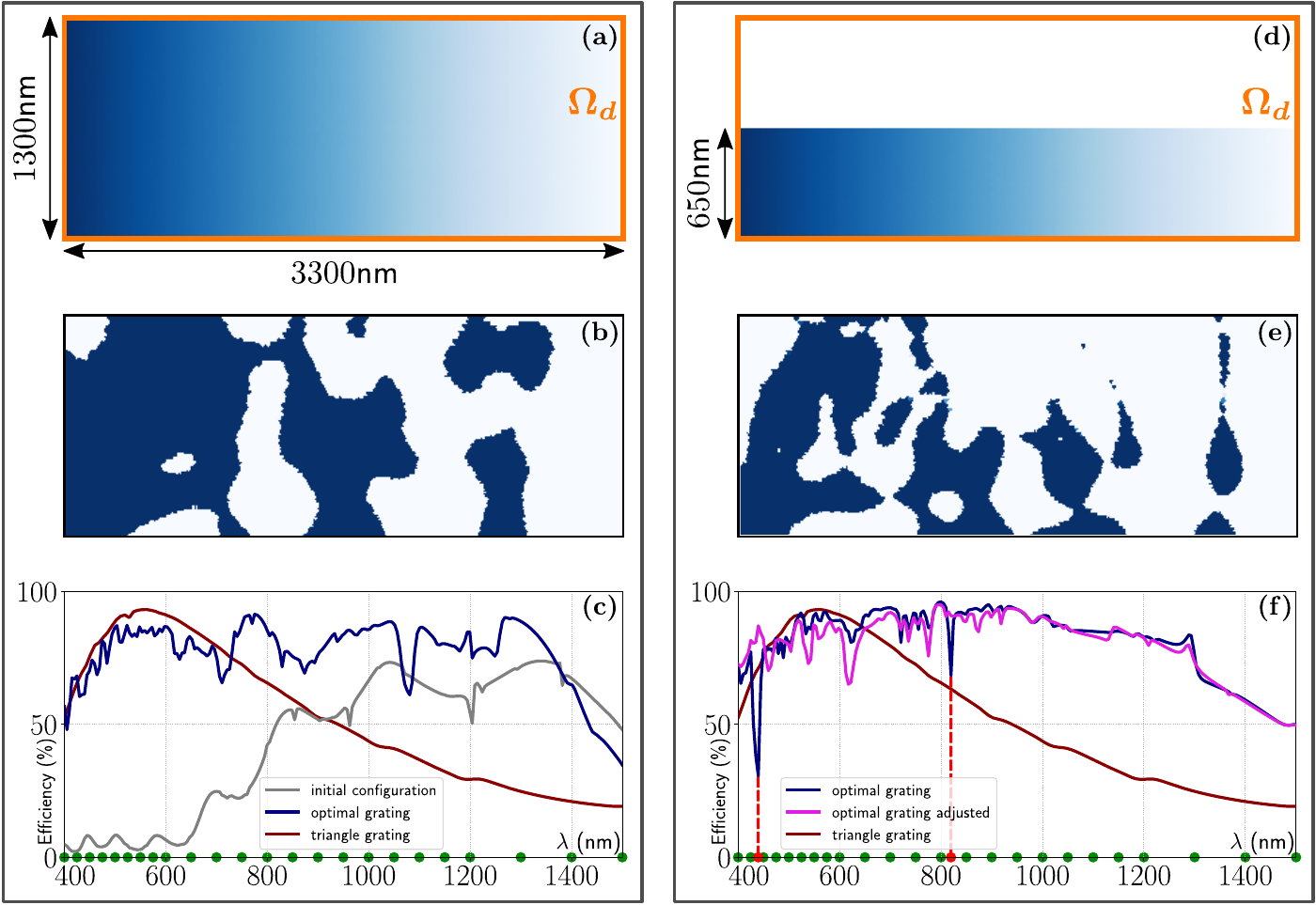}
    \caption{Results for the multi-wavelength optimization on the $-$1$^{\text{st}}$ diffraction order for a 1300\,nm-high pattern on $N_{\lambda} = 24$ target wavelengths (green dots) between 400 and 1500\,nm, for the angles $\theta_i = 5^{\circ}$, $\varphi_i = -66^{\circ}$, $\psi_i = 90^{\circ}$ with a tolerance of the target of $10^{-5}$, with connectedness (filter radius $r_f = 200$\,nm) and binarization, using two different initial configurations. \textbf{(a)}~Initial configuration with the linearly decreasing permittivity widened to the full design space. \textbf{(b)}~The corresponding optimized pattern. \textbf{(c)}~Spectral response in reflection on the [400,1500]\,nm spectral range under the target incidence angles for this optimal grating (blue curve), as compared to the initial configuration (grey curve) and the sawtooth grating (red curve). \textbf{(d)} and \textbf{(e)}~Same analysis with the initial configuration of Fig.~\ref{fig::silica_opt}\subFigRef{a} but with a design region twice higher. \textbf{(f)}~Spectral response in reflection on the [400,1500]\,nm spectral range under the target incidence angles for this optimal grating (blue curve), as compared to the sawtooth grating (red curve). Two drops are spotted in this spectral response, at 440\,nm and 820\,nm (red dots and dashed lines). Another optimization has been processed, adding these two wavelengths (violet curve).}
    \label{fig::efficiency_multiwavelength_twiceHeight}
\end{figure}

Two ways to adapt the initial configuration are chosen and lead to different optimal patterns, as displayed in Fig.~\ref{fig::efficiency_multiwavelength_twiceHeight}. The first possibility is to widen the linearly decreasing permittivity on the full design space (Fig.~\ref{fig::efficiency_multiwavelength_twiceHeight}\subFigRef{a}). The second one is to keep the same initial pattern as in Fig.~\ref{fig::silica_opt}\subFigRef{a}, while sticking to a design space that is twice higher. The rest of the design space above is then let to $\boldsymbol{\rho} = \mathbf{0}$ (Fig.~\ref{fig::efficiency_multiwavelength_twiceHeight}\subFigRef{d}).

Note that the first choice redshifts the central wavelength of the blazing effect of the initial configuration (see the grey curve in Fig.~\ref{fig::efficiency_multiwavelength_twiceHeight}\subFigRef{c}). The challenge of the optimization then consists in significantly improving the diffraction of the visible light. Running the same optimization as in the previous section on this larger design space leads to the final binary pattern shown in Fig.~\ref{fig::efficiency_multiwavelength_twiceHeight}\subFigRef{b}. It provides a convincing broadband efficiency (blue curve in Fig.~\ref{fig::efficiency_multiwavelength_twiceHeight}\subFigRef{c}) since the reflection averaged over the spectral range [400,1500]\,nm is now of 77\%, outperforming the pattern shown in Fig.~\ref{fig::efficiency_multiwavelength}\subFigRef{a} by 11\% in absolute value, and thus the classical sawtooth grating by 25\%. Therefore the relative difference between the two averaged reflectivities is 48\%, which points out an outstanding improvement.

For the initial configuration shown in Fig.~\ref{fig::efficiency_multiwavelength_twiceHeight}\subFigRef{d}, leading to the pattern of Fig.~\ref{fig::efficiency_multiwavelength_twiceHeight}\subFigRef{e}, the improvement is even more impressive, providing an averaged diffraction efficiency of 81\% over the considered bandwidth (blue curve in Fig.~\ref{fig::efficiency_multiwavelength_twiceHeight}\subFigRef{f}, as compared to the red curve of the sawtooth grating, 29\% more in absolute terms, 56\% in relative terms). However, two main drops appear at 440\,nm and 820\,nm (red dots and dashed lines in Fig.~\ref{fig::efficiency_multiwavelength_twiceHeight}\subFigRef{f}). Including these two wavelengths in the targeted-wavelength set $\Lambda$ allows to remove these drops, as highlighted by the new spectral response in violet. Other sharp drops appear on this new spectral response (for example at 620\,nm) and the efficiency averaged on the wavelength range slightly decreases to 80\%. However the global response is better distributed, which is a non negligible quality in spectroscopy applications. While the response dropped to 35\% on the first optimal pattern, it is now higher than 65\% except at the end of the interval, where the two responses are the same anyway. This kind of consideration is a way to improve the existing code, by automating the correction of deep drops in the spectral response.

Another concern occurs in the isolated elements that are small as compared to $r_f$ in Fig.~\ref{fig::efficiency_multiwavelength_twiceHeight}\subFigRef{e}. Their effect is negligible, therefore one would like to shed them. They are due to the last application of the binarization filter on the optimized pattern and illustrate the limits of the connectedness filter used here. Another way of improvement is emerging. It is linked to the wide topic of the design-rule constraints, tackled for instance in Ref.~\cite{designRule}.

\subsection{Patterning the traditional sawtooth profile}\label{sec::optimtriangular}

The model of a dielectric metasurface deposited on a metallic substrate allows to avoid the plasmonic resonances due to the sharp features of the pattern \cite{plasmonic_resonances}, a phenomenon observed on the sawtooth grating. However, at this point, one wonders what would be the outcome if the latter was considered as an initial configuration to see if the classical triangular design is actually the best metallic blazed grating.

\begin{figure}[ht!]
    \centering
    \includegraphics[width=1\textwidth]{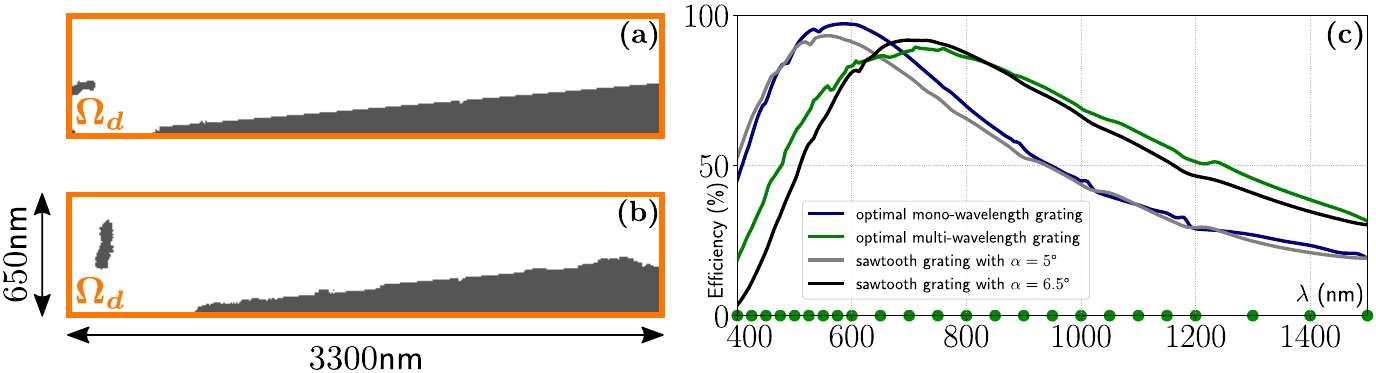}
    \caption{Results for the optimization on the $-$1$^{\text{st}}$ diffraction order for a 650\,nm-high design region for the angles $\theta_i = 5^{\circ}$, $\varphi_i = -66^{\circ}$, $\psi_i = 90^{\circ}$, a tolerance on the target of $10^{-5}$, with connectedness (filter radius of 50\,nm) and binarization. The initial configuration is a silver sawtooth pattern with a blaze angle $\alpha = 5^{\circ}$. \textbf{(a)}~Optimized pattern for a mono-wavelength optimization at $\lambda = 573$\,nm; \textbf{(b)}~Optimized pattern for a multi-wavelength optimization at the wavelengths designated by the green dots on the Fig. c; \textbf{(c)}~Spectral response in reflection on the [400,1500]\,nm spectral range under the target incidence angles for the mono-wavelength (blue curve)  and the multi-wavelength (green curve) optimized patterns, as respectively compared to the sawtooth gratings with the blaze angles $\alpha = 5^{\circ}$ (grey curve) and $\alpha = 6.5^{\circ}$ (black curve).}
    \label{fig::optim_metal}
\end{figure}

Moreover, this analysis completes the optimization process because it has been shown that a different kind of interpolation is more efficient and stable for the metallic gratings. More precisely, Christiansen \emph{et al.} have developed a non-linear interpolation scheme in Ref.~\cite{design_metal}, since the  complex refractive index $n$ is linearly interpolated instead of the relative permittivity:
\begin{equation}\label{eq::non-linear_interp}
    \begin{alignedat}{1}
        & \varepsilon_r^d(\lambda, \vec{r}) = n(\lambda, \vec{r})^2 \\
        \mbox{with } & n(\lambda, \vec{r}) = n_1(\lambda) + \big(n_2(\lambda) - n_1(\lambda)\big)\boldsymbol{\rho}(\vec{r}) ,
    \end{alignedat}
\end{equation}
$n_2$ being the refractive index of silver and $n_1$ being that of air. This new interpolation method induces a slight modification in the proposition~\ref{prop::adjoint}, detailed in the technical support, and leads to the new patterns presented in Fig.~\ref{fig::optim_metal}.

Both a mono- and a multi-wavelength optimizations have been made on the traditional silver sawtooth grating, providing resp. the patterns of Fig.~\ref{fig::optim_metal}\subFigRef{a} and~\ref{fig::optim_metal}\subFigRef{b} and the spectral responses of the blue and green curves in Fig.~\ref{fig::optim_metal}\subFigRef{c}. With the mono-wavelength optimization at $\lambda = 573$\,nm, a gain of 4\% on the maximal efficiency is observed as well as a slight shift of the maximal efficiency due to the truncation of the triangle displayed in Fig.~\ref{fig::optim_metal}\subFigRef{a}. With the multi-wavelength optimization (performed on the wavelengths highlighted by the green dots in Fig.~\ref{fig::optim_metal}\subFigRef{c}), this shift is even more pronounced. Therefore, even if the reflection averaged over the spectral range [400,1500]\,nm is of 64\% (against 52\% for the sawtooth grating with a blaze angle of 5$^{\circ}$), the lower wavelengths in the visible range are cut too much. The response obtained is actually slightly widened with respect to that of a sawtooth grating with a higher blaze angle (here $\alpha = 6.5^{\circ}$, as highlighted by the black curve). The multi-wavelength optimization has only exhibited the performances of a traditional triangular blazed grating with a maximum more centered in the [400,1500]\,nm wavelength range (here at 700\,nm).

This final study presents two advantages. On the one hand, it shows that the optimization of metallic patterns is stable and accurate with this non-linear interpolation method. On the other hand, it exhibits that the performances of the triangular profile are actually almost optimal among the metallic gratings, even by considering the non-manufacturable ones. The plasmonic resonances are consequently not the only reason why the optimization is performed on dielectrics for the blazed gratings: the triangular metallic pattern does not have much room for improvement.

\section{Conclusion}\label{sec::conclusion}

In this article, we provide all the theoretical steps allowing to compute the topology optimization with the adjoint method using the FEM under a conical incidence (2.5D). By numerically solving only two PDEs under weak formulation (a direct and an adjoint porblem), an accurate Jacobian of the reflection/transmission diffraction efficiencies with respect to a relative permittivity density is computed in order to make the optimization algorithm functional. The applications shown in the numerical section are concrete examples demonstrating the high-performance of the algorithm. The blazed diffraction efficiencies can reach 98\% for a single-wavelength optimization. Even more striking, for a multi-wavelength or broadband optimization, the traditional sawtooth grating blazed on its $-1^{\text{st}}$ diffraction order is outperformed by a topology optimized grating that increases the diffraction efficiency averaged on the [400,1500]\,nm spectral range by up to 29\% in absolute terms and 56\% in relative terms. The algorithm even points out that the usual sawtooth pattern cannot be truly improved, hence the choice to use the dielectric-on-metal model.

A lot of complementary studies are possible from now on. In the frame of open science, we provide the source code allowing to retrieve the numerical examples of this paper. Other materials, geometries, constraints and targets can be easily taken into account starting from the provided Gmsh/GetDP template model.

This paper provides theoretical and numerical grounds of a more practical ongoing work on the manufacturability of optimized blazed metasurfaces. The manufacturability of the device is a very novel field of research since the gratings already produced are either for larger wavelengths \cite{phase_shift}, or really small devices \cite{InverseDesign, Mansouree}. To our knowledge, two nano-blazed grating on a convex surface have been manufactured and tested \cite{nanoblazed_IOF,nanoblazed_TRT}. Being able to also fabricate the optimized grating would enable to oversee all the steps of the device development: theoretical modelling, manufacturing, characterization, comparison with the model and eventually integration in an actual instrument. 

\section*{Acknowledgements}
This work has been partially supported by CNES and Thales Alenia Space with a PhD grant.

\newpage
\bibliographystyle{unsrt}
\footnotesize\bibliography{bibliography}

\normalsize\section*{\hypertarget{app::appendix}{Appendix}}
\begin{appendices}
Technical proofs and validations that are not essential for the understanding of the paper are detailed here. The same notations and definitions as in the article are used.

First, additional information is given on the connectedness and binarization filters on the density field and their Jacobian in appendix~\ref{app::filters}. Then the proof of proposition~\ref{prop::adjoint} and its corollary are given in appendix~\ref{app::proof_prop2}, which can be useful to the curious reader who needs to modify the cost function. A validation of the computation of the Jacobian through this adjoint method is shown afterwards in appendix~\ref{app::validation}, by direct comparison to a brute element-by-element finite difference approach. Finally, the choice of the initial configuration of the optimization is explained in appendix~\ref{app::initial_config}.

\section{Filters applied and their Jacobian}\label{app::filters}

The diffusion filter used here is a linear function that homogenizes the values around every mesh element with a given radius $r_f$ \cite{TD_3D_TopOpt}. Let $\mathcal{T}_i$ be the triangle with the density $\rho_i$. Let then $\text{dist}_{j,r_f}$ be a function such that
\begin{equation}\label{eq::def_dist_rf}
    \text{dist}_{j,r_f}(\rho_i) =
    \left\{
        \begin{array}{llc}
            \text{dist}(\mathcal{T}_i,\mathcal{T}_j) & \mbox{if the latter is smaller than $r_f$}\\
            0                                        & \mbox{else}
        \end{array}
    \right.,
\end{equation}
where the distance between two triangles is the distance between their barycenters. This distance function selects the surroundings triangles closer than the filter radius. Then the diffusion filter function is defined by: 
\begin{equation}\label{eq::def_diffusion_filter}
    \rho_{f,i}(\boldsymbol{\rho}) = \frac{\sum_{j=1}^N \text{dist}_{j,r_f}(\rho_i) \rho_j}{\sum_{j=1}^N \text{dist}_{j,r_f}(\rho_i)}.
\end{equation}
In essence, a sliding averaging is performed on the densities. One can actually recognize a linear matrix function. Let $\mathbb{D}$ be the matrix of the different values of $\text{dist}_{j,r_f}$: $\mathbb{D}_{i,j} = \text{dist}_{j,r_f}(\rho_i)$ and let $\mathfrak{1}$ be the column vector with ones. Then we can rewrite the filter function
\begin{equation}\label{eq::def_diffusion_filter_mat}
    \boldsymbol{\rho}_f(\boldsymbol{\rho}) = \frac{\mathbb{D}\boldsymbol{\rho}}{\mathbb{D}\mathfrak{1}}
\end{equation}
where the division between vectors in the last expression designates the division element per element. This expression directly leads to the Jacobian of the filter:
\begin{equation}\label{eq::jac_diffusion_filter_mat}
    \frac{\partial \boldsymbol{\rho}_f}{\partial \boldsymbol{\rho}}(\boldsymbol{\rho}) = \frac{\mathbb{D}}{\mathbb{D}\mathfrak{1}}
\end{equation}
where the division between a matrix and a vector in the last expression is the division between each line of the matrix by this vector.

Concerning the binarization filter, a mere analytic differentiation leads to
\begin{equation}\label{eq::derivative_binarization}
    \frac{\partial \hat{\rho}_i}{\partial \rho_{f,j}}(\boldsymbol{\rho}_f)
    = \frac{\beta_f (1-\tanh^2[\beta_f(\rho_{f,i} - \nu)])}{\tanh(\beta_f\nu) + \tanh[\beta_f(1 - \nu)]} \delta_{ij},
\end{equation}
where $\delta_{ij}$ is the Kronecker delta. The matrix multiplication \eqref{eq::jac_diffusion_filter_mat}$ \times $\eqref{eq::derivative_binarization} (in this order) provides the filtering factor in the derivative of $\mathcal{F}_n$. The last factor for the total derivative is given by the proposition~\ref{prop::adjoint}, which proof is detailed in the next section of this technical support.
\newpage
\section{Proof of the proposition~\ref{prop::adjoint} and its corollary}\label{app::proof_prop2}

\subsection{Proof of the proposition}

This proof is adapted from the general result presented in Ref.~\cite{mixedOpt_Geuzaine}. For the sake of simplicity, we denote only by $\boldsymbol{\rho}$ the filtered density field (instead of $\boldsymbol{\hat{\rho}}_f$). First it is important to notice that $r_n^u$ depends explicitly on $\vec{E}_*^d$ that depends implicitly on $\boldsymbol{\rho}$. The derivative of the composite function writes then for all $\boldsymbol{\rho} \in \mathbb{R}^N$:
\begin{equation}\label{eq::derivative_rn}
    \frac{\partial r_n^u}{\partial \rho_i}(\vec{E}_*^d(\boldsymbol{\rho}))
    = \restriction{\frac{\partial r_n^u}{\partial \rho_i}}{\frac{\partial\vec{E}}{\partial  \rho_i} = 0} (\vec{E}_*^d(\boldsymbol{\rho}))
        + \Bigl\{ \mathrm{D}_{\vec{E}} r_n^u \Bigr\} \biggl( \frac{\partial \vec{E}_*^d}{\partial \rho_i}(\boldsymbol{\rho}) \biggr),
\end{equation}
where $\mathrm{D}_{\vec{E}}$ designates the Fréchet derivative which is the linear operator such that \cite{Frechet_diff}:
\begin{equation}\label{eq::def_Frechet}
    r_n^u(\vec{E}_*^d(\boldsymbol{\rho}) + \delta\vec{E}(\boldsymbol{\rho})) = r_n^u(\vec{E}_*^d(\boldsymbol{\rho})) + \Bigl\{ \mathrm{D}_{\vec{E}}r_n^u \Bigr\} (\delta\vec{E}(\boldsymbol{\rho})) + o(\delta\vec{E})
\end{equation}
with $$\lim_{\lVert \delta\vec{E} \rVert \to 0} \frac{1}{\lVert \delta\vec{E} \rVert} o(\delta\vec{E}) = 0.$$ As $r_n^u$ is linear with respect to $\vec{E}_*^d(\boldsymbol{\rho})$, the first term in \eqref{eq::derivative_rn} vanishes and moreover
\begin{align*}
    r_n^u(\vec{E}_*^d(\boldsymbol{\rho}) + \delta\vec{E}(\boldsymbol{\rho}))
    & = \frac{1}{d}\int_0^d e^{-i\alpha_n x}(\vec{E}_*^d(\boldsymbol{\rho}) + \delta\vec{E}(\boldsymbol{\rho}) + \vec{E}_a^d) \cdot \hat{\vec{u}} \, \mathrm{d}x \\
    & = \frac{1}{d}\int_0^d e^{-i\alpha_n x}(\vec{E}_*^d(\boldsymbol{\rho}) + \vec{E}_a^d) \cdot \hat{\vec{u}} \, \mathrm{d}x + \frac{1}{d}\int_0^d e^{-i\alpha_n x} \delta\vec{E}(\boldsymbol{\rho}) \cdot \hat{\vec{u}} \, \mathrm{d}x.
\end{align*}
Thus we deduce that for all $\boldsymbol{\rho} \in \mathbb{R}^N$:
\begin{equation}\label{eq::Frechet_rn}
    \Bigl\{ \mathrm{D}_{\vec{E}}r_n^u \Bigr\} \biggl( \frac{\partial \vec{E}_*^d}{\partial \rho_i} (\boldsymbol{\rho}) \biggr)
    = \frac{1}{d} \int_0^d e^{-i\alpha_n x}\frac{\partial \vec{E}_*^d}{\partial \rho_i} (\boldsymbol{\rho}) \cdot \hat{\vec{u}} \, \mathrm{d}x
    = \frac{\partial r_n^u}{\partial \rho_i}(\vec{E}_*^d(\boldsymbol{\rho})).
\end{equation}
This derivative cannot be calculated numerically since it would rely on a computation of every derivative of $\partial \vec{E}_*^d/\partial \boldsymbol{\rho}$. The same consideration as for $r_n^u$ applies here: one could use numerical finite differences. However an alternative method is to consider the augmented Lagrangian of $r_n^u$ regarding the Helmholtz PDE.

For $u=\{x,z\}$, let $r_n^u$ be a performance function and $r_{n,a}^u$ its so-called augmented Lagrangian:
\begin{equation}\label{eq::rn_aug_Lag}
    r_{n,a}^u(\vec{E}_*^d(\boldsymbol{\rho}),\vec{E}') = r_n^u(\vec{E}_*^d(\boldsymbol{\rho})) - \mathcal{L}(\boldsymbol{\rho},\vec{E}').
\end{equation}
Then the same reansoning as equations \eqref{eq::derivative_rn} to \eqref{eq::Frechet_rn} is made for $\mathcal{L}$ and the derivatives of this object are for all $(\boldsymbol{\rho},\boldsymbol{\lambda}) \in \mathbb{R}^N \times \boldsymbol{\mathcal{V}}$,
\begin{equation}\label{eq::deriv_rn_aug_Lag}
    \begin{alignedat}{1}
        \frac{\partial r_{n,a}^u}{\partial\rho_i}(\vec{E}_*^d(\boldsymbol{\rho}),\vec{E}')
        = & \int_0^d \frac{1}{d} e^{-i\alpha_n x}\frac{\partial \vec{E}_*^d}{\partial \rho_i} (\boldsymbol{\rho}) \cdot \hat{\vec{u}} \, \mathrm{d}x
            - \int_{\Omega} \Biggl[ \tens{\mu_r}^{-1} \rot\frac{\partial \vec{E}_*^d}{\partial \rho_i} (\boldsymbol{\rho}) \cdot \rot\overline{\vec{E}'} \\
          & - k_0^2 \biggl( \tens{\varepsilon_r}\frac{\partial \vec{E}_*^d}{\partial \rho_i}(\boldsymbol{\rho})
                + \frac{\partial \tens{\varepsilon_r}}{\partial \rho_i}(\boldsymbol{\rho})\underbrace{(\vec{E}_*^d(\boldsymbol{\rho}) + \vec{E}_a)}
                                _{\vec{E}^{\text{tot}}}
            \biggr) \cdot \overline{\vec{E}'} \Biggr] \, \mathrm{d}\Omega .
    \end{alignedat}
\end{equation}
We consider now that, for a given $i\in\{ 1,...,N\}$, $\boldsymbol{\lambda} := \overline{\vec{E}'}$ is a function of an adjoint functional space $\boldsymbol{\mathcal{V}}^{\text{adj}}$ that is to be characterized and $\boldsymbol{\lambda}' := \partial \overline{\vec{E}_*^d} / \partial\rho_i $ is a test function in this same adjoint space. The crucial remark to understand the adjoint method is that, if there exists a function $\boldsymbol{\lambda}_*^u = \overline{\vec{E}'}$ such that only the last term in \eqref{eq::deriv_rn_aug_Lag} (with the underbrace) remains, then it would be possible to compute the derivatives of the augmented Lagrangian. Furthermore, for the equilibrium design variable $\boldsymbol{\rho}_*$ such that $\mathcal{L}(\boldsymbol{\rho_*},\vec{E}') = 0$ for all $\vec{E}' \in \boldsymbol{\mathcal{V}}$, we can state \cite{ShapeOpt_Geuzaine} that $\frac{\partial \mathcal{L}}{\partial\rho_i} (\boldsymbol{\rho}_*)= 0$, so that
\begin{equation}\label{eq::equality_rn_aug_Lag}
    \frac{\partial r_{n,a}^u}{\partial\rho_i}(\vec{E}_*^d(\boldsymbol{\rho}_*),\boldsymbol{\lambda}_*^u)
    = \frac{\partial r_n^u}{\partial\rho_i}(\vec{E}_*^d(\boldsymbol{\rho}_*)).
\end{equation}
In other terms, finding such an \textit{adjoint variable} $\boldsymbol{\lambda}_*^u$ would solve the issue of calculating the derivatives of $r_n^u$. Let then consider the so-called adjoint problem:
\medbreak
\noindent Find
$\boldsymbol{\lambda}_*^u\in\boldsymbol{\mathcal{V}}^{\text{adj}}$ such that for all $\boldsymbol{\lambda}' \in \boldsymbol{\mathcal{V}}^{\text{adj}}$,
\begin{equation}\label{eq::adjoint_problem_appendix}
    \int_{\Omega}
    \Biggl[
    \tens{\mu_r}^{-1} \rot\boldsymbol{\lambda}_*^u \cdot \rot\overline{\boldsymbol{\lambda}'}
    - k_0^2 \tens{\varepsilon_r}\boldsymbol{\lambda}_*^u \cdot \overline{\boldsymbol{\lambda}'}
    \Biggr]\,\mathrm{d}\Omega
    - \frac{1}{d} \int_0^d e^{-i\alpha_n x} \hat{\vec{u}} \cdot\overline{\boldsymbol{\lambda}'} \,\mathrm{d}x
    = 0.
\end{equation}
This problem has a unique solution indeed, since it is the weak formulation of a Maxwell's equation with a surface current $\vec{j} := e^{-i\alpha_n x} / d \,\, \hat{\vec{u}}$ as a source.

Eventually, combining ~\eqref{eq::equality_rn_aug_Lag} with the injection of the solution of ~\eqref{eq::adjoint_problem_appendix} into ~\eqref{eq::deriv_rn_aug_Lag} provides all the derivatives of $r_n^u$ with respect to the $\rho_i$ around the equilibrium point $\boldsymbol{\rho}_*$:
\begin{equation}\label{eq::adjoint_derivative_rn_appendix}
    \frac{\partial r_n^u}{\partial \rho_i} (\boldsymbol{\rho}_*)
    = \int_{\Omega} k_0^2 \frac{\partial \tens{\varepsilon_r}}{\partial \rho_i}(\boldsymbol{\rho}) \vec{E}^{\text{tot}} (\boldsymbol{\rho}_*) \cdot \boldsymbol{\lambda}_*^u \, \mathrm{d}\Omega.
\end{equation}
The last step is to express the derivatives of $\tens{\varepsilon}_r$ with respect to $\rho_i$. The relative permittivity actually varies only in the design region. Moreover, a variation around a single density $\rho_i$ only induces a variation in the triangle $\mathcal{T}_i$. Lastly, since the interpolation methods seen in the article provide a direct link between $\varepsilon_r^d$ and $\boldsymbol{\rho}$, the expression of $\partial \tens{\varepsilon}_r / \partial \rho_i$ is:
\begin{itemize}
    \item $\varepsilon_{r,\text{diel}} - \varepsilon_r^+$ in a dielectric, using the linear SIMP interpolation method \cite{SIMP_sigmund}, which concludes the proof of the proposition \ref{prop::adjoint} ;
    \item in a metal, using the non-linear interpolation method \cite{design_metal},
    \begin{equation*}
        \frac{\partial \tens{\varepsilon_r}}{\partial \rho_i}(\boldsymbol{\rho})
        = 2(n_2 - n_1) \cdot \big(n_1 + \boldsymbol{\rho}(n_2 - n_1)\big).
    \end{equation*}
\end{itemize}

\subsection{Proof of the corollary}

As it can be seen between the corollaries~\ref{cor::direct_problem} and \ref{cor::adjoint_problem}, the only non-trivial differences between both weak formulations are the signs before the $i\gamma$ terms. This aspect is detailed here. More precisely, the central element of this proof is the characterization of $\boldsymbol{\mathcal{V}} ^{\text{adj}}$. Let $\vec{E}$ be a function of $\boldsymbol{\mathcal{V}}$. In the conical case, it means that it can be decomposed using its quasi-periodicity with a factor $\alpha$ on the one hand and the exponential dependency in $z$ on the other hand. Then for all $\vec{x} := (x,y,z) \in \mathbb{R}^3$,
\begin{equation}\label{eq::decomp_elmt_V}
    \vec{E}(\vec{x}) = \vec{E}^\#(x,y) e^{i\alpha x} e^{i\gamma z}
\end{equation}
with $\vec{E}^\#$ a field periodic along $x$. Also by definition $\overline{\vec{E}} \in \boldsymbol{\mathcal{V}}^{\text{adj}}$ and we have that
\begin{equation}\label{eq::conj_elmt_V}
    \overline{\vec{E}}(\vec{x}) = \overline{\vec{E}^\#}(x,y) e^{-i\alpha x} e^{-i\gamma z}.
\end{equation}

The reciprocal of this statement is immediate by using the exact same arguments. The adjoint space therefore writes $\boldsymbol{\mathcal{V}}^{\text{adj}} = \boldsymbol{\mathcal{H}}_0(\rot_{-\gamma}, \Omega, e^{-i\alpha d})$. Furthermore, using now $\rot_{-\gamma}$ for the elements of $\boldsymbol{\mathcal{V}}^{\text{adj}}$ leads to the sign change in the weak formulation of the corollary~\ref{cor::adjoint_problem}.
\newpage
\section{Numerical validation of the adjoint method}\label{app::validation}

The Finite Element Method for the direct problem or the convergence of the methods are common elements that have to be checked in order to certify that the algorithm provides a trustable result. An important point here is to check the computation of the Jacobian with the adjoint method. To do so, we compare it to the Jacobian found with finite differences.

\begin{figure}[ht!]
    \centering
    \includegraphics[width=1\textwidth]{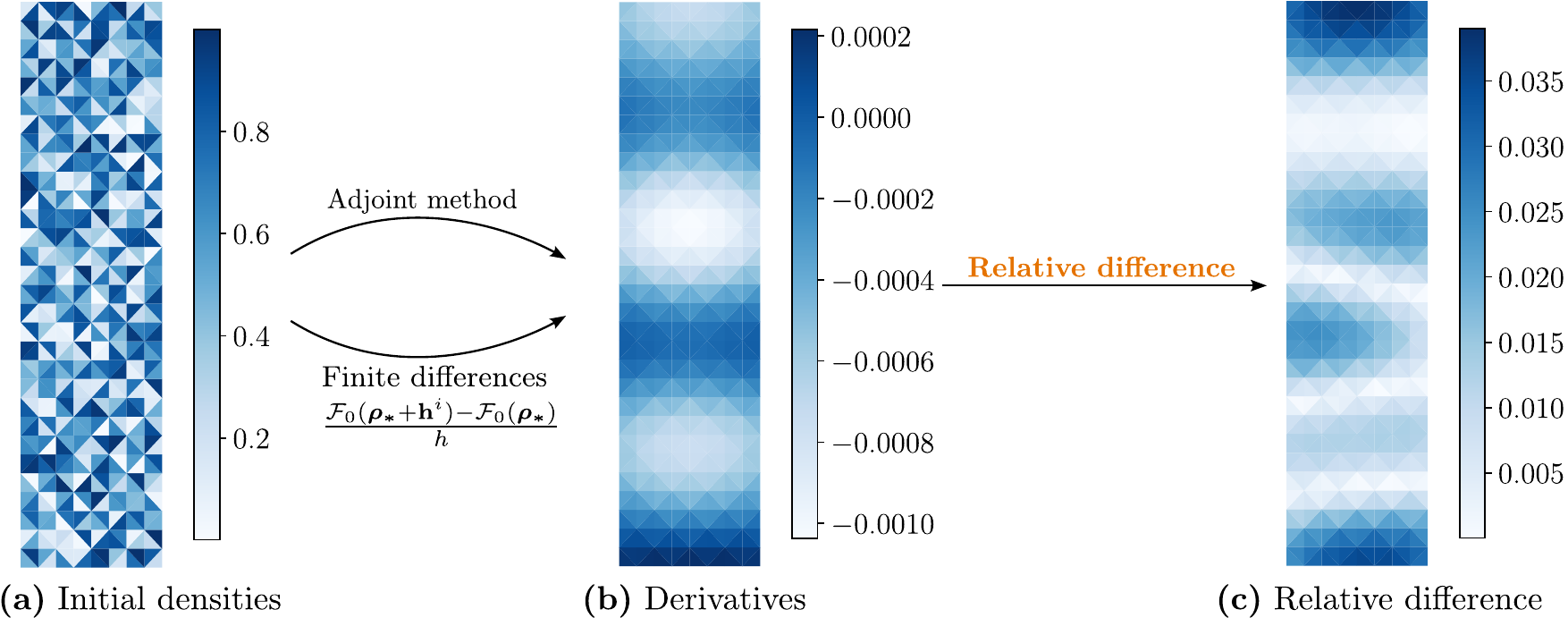}
    \caption{Test of the adjoint method on a small grid for an incident field with angles $\theta_i = 5^{\circ}$, $\varphi_i = -66^{\circ}$, $\psi_i = 90^{\circ}$ and with a multi-wavelength target function with wavelengths $\{400,700,900\}$\,nm on the specular order. \textbf{(a)}~Initial random distribution of densities. \textbf{(b)} Derivatives obtained for each mesh element, both lead to the same distribution. \textbf{(c)} Relative differences between both methods (adjoint and finite differences) for each mesh element.}
    \label{fig::test_adjoint}
\end{figure}

Let consider a grid with a random density distribution (see Fig.~\ref{fig::test_adjoint}\subFigRef{a}). The derivative of the target is computed both ways, as illustrated on Fig.~\ref{fig::test_adjoint}: first with the adjoint method and then with the finite differences. The direct problem is solved $N$ times in order to have the variation of $\mathcal{F}_n$ when adding a small step $h$ to $\rho_i$ for every $i$ (for every triangle). Note that the derivatives of the filters described in section 2 of this document are included in this study. For this test, the step of the finite differences is equal to $h=10^{-2}$ and the binarization factor is $\beta_f = 8$. Also, a conical incidence is chosen, with $\theta_i = 5^{\circ}$, $\varphi_i = -66^{\circ}$ and $\psi_i = 90^{\circ}$, just like in the optimization case of the article. Finally, this validation is made for a multi-wavelength target function with wavelengths $\{400,700,900\}$\,nm. The relative differences between the adjoint method and finite differences displayed on Fig.~\ref{fig::test_adjoint}\subFigRef{c} are defined on each mesh element $\mathcal{T}_i$ by:
\begin{equation}\label{eq::relative_diff}
    d_{a,f} = \left|\frac{\partial_i\mathcal{F}_{\text{finiteDiff}} - \partial_i\mathcal{F}_{\text{adjoint}}}{\text{mean}|\partial_i\mathcal{F}_{\text{finiteDiff}}|}\right|.
\end{equation}
It is smaller than $4\%$ in all mesh elements, even with this coarse mesh. This error decreases when the mesh is refined.

\section{Choice of the initial configuration}\label{app::initial_config}

\begin{figure}[ht!]
    \centering
    \includegraphics[width=1\textwidth]{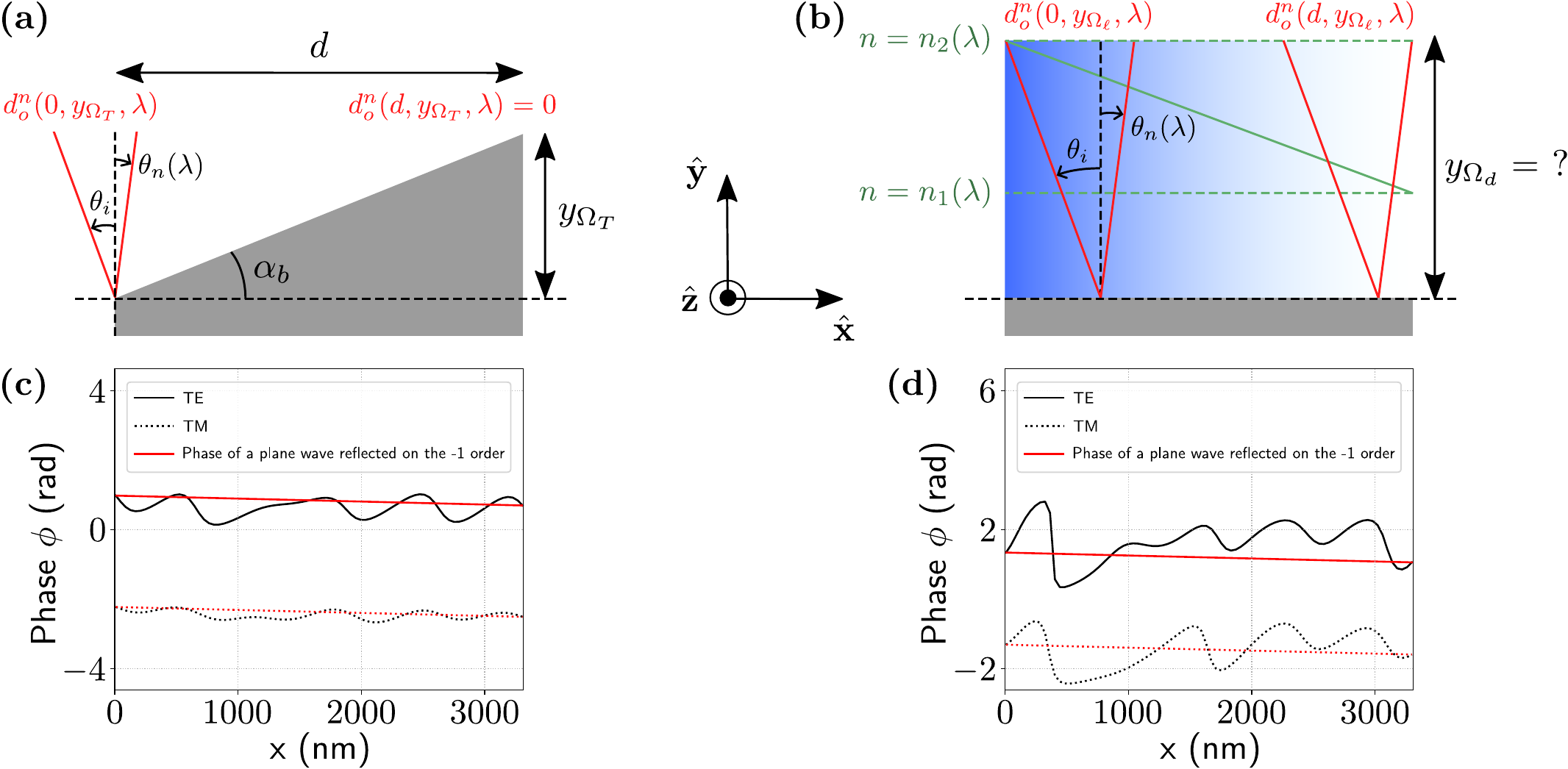}
    \caption{Imitation of the phase shift $\Delta\phi$ induced by a triangular grating providing the initial configuration for the optimization, described with the optical paths of the light in both configurations. \textbf{(a)} Optical path for the triangular grating. \textbf{(b)} Approximated optical path for the equivalent density distributed pattern. \textbf{(c)} (black lines) Phase of the diffracted field on a period of the triangular pattern ($x\in$[0,3300]\,nm) with a 2D incoming field with $\lambda=600$\,nm, $\theta_i=5^{\circ}$ in TE (solid lines) and TM (dashed lines) polarizations. (red lines) Ideal case where a plane wave is reflected on the $-1^{\text{st}}$ diffraction order. \textbf{(d}) Same analysis for the equivalent density distribution pattern.}
    \label{fig::phase_shift}
\end{figure}

For any optimization process, an initial configuration is required. Its choice is not obvious at first sight and it is important since it conditions the minimum of the cost function found by the optimization algorithm. In this study, we decided to consider as initial configuration blazed grating made of a graded indexed dielectric lying on a silver substrate. The linear decay of the graded permittivity (see Fig~\ref{fig::phase_shift}b) in the design region tries to mimic the phase shift obtained with a geometrical linear ramp of the silver triangle blazed grating (see Fig~\ref{fig::phase_shift}a). The only unknown in this problem is the height $y_{\Omega_d}$ of the graded layer. It is found by considering the difference of optical paths between $x=$ 0\,nm and $x=$ 3300\,nm as shown in red color in Fig.~\ref{fig::phase_shift}\subFigRef{a} and \subFigRef{b}, similarly to the (approximate) rule of thumb used for traditional blazed gratings.

In Fig.~\ref{fig::phase_shift}\subFigRef{a}, the height $y_{\Omega_T}$ of the silver triangle is $d\tan \alpha$. Therefore the difference $\Delta d_{o,T}^n (\lambda)$ of optical paths between $x=$ 0\,nm and $x=$ 3300\,nm for the triangle is
\begin{equation}\label{eq::don_triangle}
    \Delta d_{o,T}^n (\lambda) = -d\tan \alpha \Big(\frac{1}{\cos\theta_i} + \frac{1}{\cos\theta_n (\lambda)}\Big).
\end{equation}
For the graded-indexed grating shown in Fig.~\ref{fig::phase_shift}\subFigRef{b}, the optical rays of interest are the ray entering the cell at $x=0$ and the ray leaving the cell at $x=d$.
The horizontal size of the region where those rays and their respective reflection travel is $s_h(\lambda) = d|\sin\theta_i - \sin\theta_n(\lambda)|$. Let now consider that $\theta_i \ll \pi/2$ and $\theta_n (\lambda) \ll\pi/2$. In that case $s_h(\lambda) \ll d$ and it can be considered that the first ray evolves in a dielectric with $n = n_2(\lambda)$ and the second one with $n = n_1(\lambda)$. This is why approximately, the optical distance for the first light beam is $n_2(\lambda) y_{\Omega_d}(\lambda)(1/\cos\theta_i + 1/\cos\theta_n(\lambda))$, and $n_1(\lambda) y_{\Omega_d}(\lambda)(1/\cos\theta_i + 1/\cos\theta_n(\lambda))$ for the second one. The optical path difference $\Delta d_{o,\ell}^n (\lambda)$ between the two ends of the design region is thus:
\begin{equation}\label{eq::don_design}
    \Delta d_{o,\ell}^n (\lambda) = y_{\Omega_d}(\lambda) (n_1(\lambda) - n_2(\lambda)) \Big(\frac{1}{\cos\theta_i} + \frac{1}{\cos\theta_n(\lambda)}\Big).
\end{equation}
Setting this optical path difference  $\Delta d_{o,\ell}^n (\lambda)$ to that of the traditional triangular grating $ \Delta d_{o,T}^n (\lambda)$ leads to:
\begin{equation}\label{eq::value_yomegal}
    y_{\Omega_d}(\lambda) = \frac{d\tan\alpha}{n_2(\lambda) - n_1(\lambda)}.
\end{equation}
The higher $n_2 (\lambda)$ is, the lower $y_{\Omega_d} (\lambda)$ is. For the multi-wavelength optimization, the range [400,1500]nm is considered and the lower refractive index of silica is for $\lambda = $ 1500\,nm. In this case $n_2(\lambda) = 1.445$ and by taking $d = $ 3300\,nm and $\alpha = 5^{\circ}$, we obtain $y_{\Omega_d}(\lambda) \simeq $ 650\,nm.

An illustration of the phase shift induced by the graded-indexed structure is shown in Fig.~\ref{fig::phase_shift}\subFigRef{d} and can be compared to the phase shift induced by the triangular grating shown in Fig.~\ref{fig::phase_shift}\subFigRef{c}. The phase of the diffracted field computed using the FEM is plotted in black lines for an incoming field with a small angle of incidence ($5^{\circ} \simeq 0.087\,\text{rad}\ll \pi/2\,\text{rad}$). The wavevector lies in the plane of incidence ($\varphi_i=0$) and the computation is made in TE (solid lines) and TM (dashed lines) polarizations cases. We compare the phase shift obtained numerically to a perfect plane wave reflected on the $-$1$^{\text{st}}$ diffraction order with a $y$-intercept chosen the same as the for real field (red lines). It shows that the approximate design rule guided by physical optics provides a satisfying phase shift. Actually, the spectral response of this initial configuration (in grey on Fig.~\ref{fig::efficiency_monowavelength}\subFigRef{a} of the article for instance) indicates that this grating is blazed for a quite broad wavelength range, the blazed efficiency being over 50\% in the [450,1100]\,nm range.
\end{appendices}

\end{document}